\documentclass[a4paper,11pt]{article}
\pdfoutput=1 

\usepackage{jcappub} 

\usepackage[T1]{fontenc} 

\usepackage{graphicx}

\title{Superradiant Bose--Einstein condensates around Kerr black holes}

\renewcommand{\thefootnote}{\fnsymbol{footnote}}
\author[a]{Sen Guo}
\author[a,*]{Lin Wen}
\author[a,*]{Xiao-Xiong Zeng}

\affiliation[a]{College of Physics and Electronic Engineering, Chongqing Normal University, Chongqing 401331, People's Republic of China}

\renewcommand{\thefootnote}{\fnsymbol{footnote}}
\footnotetext[1]{Corresponding author.}
\renewcommand{\thefootnote}{\arabic{footnote}}

\emailAdd{sguophys@126.com}
\emailAdd{wlqx@cqnu.edu.cn}
\emailAdd{xxzengphysics@163.com}

\abstract{Ultralight bosonic dark matter can accumulate around a rotating black hole, where superradiance amplifies the field until a macroscopic cloud forms. Whether such a cloud behaves as a Bose--Einstein condensate depends on the self-interaction, which earlier work has either retained on static backgrounds or dropped on the Kerr metric. Here we treat the two together. Starting from the Klein--Gordon equation with a quartic potential, we separate the linear problem into spheroidal and radial equations and solve them self-consistently, obtain the superradiant growth rate from the conserved Noether current, project the nonlinear term onto a single mode, and integrate the resulting Gross--Pitaevskii equation with a bordered Newton method at fixed particle number. Rotation modulates the self-interaction geometrically: the effective coupling carries a factor $\Delta(r)$ and therefore switches off at the horizon. Once the field is rescaled, the whole solution family depends on the single dimensionless parameter $\mathcal{N}=\lambda N$. We recover the hydrogenic spectrum of the gravitational atom and the $\alpha^{4\ell+5}$ scaling of the growth rate, and we obtain the exact relation $J_{z}=\hbar mN$, which receives no correction from the self-interaction. We find that the condensate is a torus rather than a spherical shell, with its density vanishing identically on the rotation axis, and that the cloud is modified appreciably only for $\mathcal{N}\gtrsim10^{3}$. We also find that the repulsion can quench superradiance only inside a window of relative width $\alpha^{2}/2n_{p}^{2}$, about one per cent, so the self-interaction cannot saturate the instability by itself and the black hole must spin down instead. For a $10M_{\odot}$ host at $\alpha=0.1$ the emission falls near $650\,\mathrm{Hz}$. The toroidal geometry has no counterpart in spherical symmetry and should be taken into account when the gravitational-wave signal from these clouds is modelled.}

\begin{document}
\maketitle
\flushbottom
\renewcommand{\thefootnote}{\arabic{footnote}}

\section{Introduction}
\label{sec:intro}

\par
The flattening of galactic rotation curves, the offset between the lensing mass and the baryonic mass in cluster collisions, and the anisotropy spectrum of the cosmic microwave background all point to a matter component that manifests itself only through gravity, and whose energy density accounts for about a quarter of the total energy density of the Universe~\cite{Rubin:1970zza,Clowe:2006eq,Planck:2018vyg}. The standard cold dark matter picture is remarkably successful on large scales. On small scales, however, several long-standing tensions persist, among them the core--cusp problem of galactic density profiles, the discrepancy between the predicted and the observed abundance of satellite galaxies, and the so-called ``too-big-to-fail'' problem~\cite{deBlok:2009sp,Moore:1999nt,Boylan-Kolchin:2011qkt}. All of these tensions arise on the very scales at which cold dark matter phenomenology relies most heavily on the assumption that the particle mass is large enough for the de Broglie wavelength to be negligible. If dark matter consists instead of extremely light bosons, structure formation is suppressed on small scales by quantum pressure~\cite{Hu:2000ke}. For a scalar particle of mass $M_{\rm boson}c^{2}\sim10^{-22}\,\mathrm{eV}$, the de Broglie wavelength is of order a kiloparsec, which is comparable to the size of a galactic core~\cite{Schive:2014dra,Hui:2016ltb}. The cosmological and astrophysical constraints on this scenario are reviewed in Refs.~\cite{Marsh:2015xka,Ferreira:2020fam}. From the particle-physics side, the Peccei--Quinn axion introduced to solve the strong CP problem is itself a pseudo Nambu--Goldstone boson, and string compactifications generically produce a large number of axion-like particles whose masses span many orders of magnitude~\cite{Wilczek:1977pj,Arvanitaki:2009fg}. An ultralight scalar field is therefore motivated both cosmologically and from particle physics.

\par
The occupation number of ultralight bosons is enormous, so that they may be described by a classical field. When they occupy a single macroscopic quantum state, the appropriate description is a Bose--Einstein condensate. The theory of the relativistic Bose gas provides the basis for such a description, and a series of subsequent works has explored the possibility of modelling dark matter as a scalar-field condensate, together with the halo structure and the cosmological evolution that follow~\cite{Parker:1991jg,Sin:1992bg,Ji:1994xh,Matos:1998vk,Boehmer:2007um,Urena-Lopez:2008vpl,Suarez:2013iw}. A condensate must, however, be distinguished from a free massive scalar cloud, the difference between them being the self-interaction. The density profile, the chemical potential and the Thomas--Fermi radius of a condensate all originate in the self-interaction; without it the Gross--Pitaevskii equation reduces to a linear Schr\"odinger equation, and little of the notion of a condensate
survives~\cite{Chavanis:2011zi}. In a relativistic framework the corresponding step is to add a self-interaction potential $V(\Phi)=\mu^{2}|\Phi|^{2}+\frac{\lambda}{2}(|\Phi|^{2})^{2}$ to the Klein--Gordon equation. The resulting equation is structurally similar to the non-relativistic Gross--Pitaevskii equation and is accordingly referred to as a Gross--Pitaevskii-type equation~\cite{Chavanis:2011zi}. The present work is organized around the role of $\lambda$, since it is precisely this coupling that turns a scalar cloud into a condensate.

\par
Supermassive black holes reside at the centres of most galaxies, and the imaging of M87\* and Sgr~A\* by the Event Horizon Telescope has turned this inference into a direct observation~\cite{EventHorizonTelescope:2019dse,EventHorizonTelescope:2022wkp}. If dark matter is an ultralight scalar field, how is it distributed near such a black hole? For the Schwarzschild spacetime, Barranco \textit{et al.} showed that a massive scalar field without self-interaction admits quasibound states, that is, spherically symmetric configurations with no ingoing flux either at infinity or at the horizon~\cite{Barranco:2012qs}. Such configurations must decay in time, because particles tunnel through the barrier into the horizon; their decay time can nevertheless be as long as the age of the Universe, and they have therefore been termed scalar wigs~\cite{Barranco:2013rua}. This result was later extended to the self-gravitating case, in which the existence of quasistationary clouds was confirmed by solving the coupled Einstein--Klein--Gordon system numerically~\cite{Barranco:2017aes,Sanchis-Gual:2014ewa}. The self-interaction was incorporated by Castellanos \textit{et al.}, who derived
Gross--Pitaevskii-type equations on the Schwarzschild and Schwarzschild--de~Sitter backgrounds, analysed the trapping condition of the effective potential, and tested the applicability of the Thomas--Fermi approximation~\cite{Castellanos:2013ena,Castellanos:2015nbe}. The most recent study of that series turned to charged black holes, and examined three spherically symmetric static solutions arising from different electrodynamic theories: the Reissner--Nordstr\"om solution of Maxwell theory, the Hoffmann solution of Born--Infeld theory, and the regular Ay\'on-Beato--Garc\'ia solution of a nonlinear electrodynamics~\cite{Castellanos:2017tka}. The three share the same asymptotics outside the horizon while differing markedly in their interior structure, which raises a natural question: can a condensate cloud outside the horizon be used as a probe to distinguish them from the exterior? The answer obtained in Ref.~\cite{Castellanos:2017tka} is negative. Even for the largest charge parameter at which the three can be compared, their condensate density profiles almost coincide, and the differences between them are smaller than the error of the Thomas--Fermi approximation itself. Two conclusions of that work bear directly on the present one. First, the existence of a local minimum of the effective potential, and hence the possibility of partial trapping, requires $\mu$ to be sufficiently small; for a supermassive black hole this gives $M_{\rm boson}c^{2}\lesssim10^{-14}\,\mathrm{eV}$. Second, for the parameters adopted there, a cloud around a black hole of the M87\* scale lives for only about $3.5\times10^{4}$ years and carries an energy of order $10^{-60}$ of the black-hole mass. The picture obtained in the spherically symmetric case is therefore observationally pessimistic, for three reasons. The barrier is of finite height, so that particles tunnel continuously into the horizon, the net number flux is always directed inwards, and the cloud can only decay. A lifetime
of astrophysical interest requires the frequency $\omega$ to be tuned extremely close to $\mu$. The gravitational effect of the cloud, finally, lies far below any foreseeable detection threshold. It should be stressed that these three points are not evidence against scalar-field dark matter, but consequences of the assumption of spherical symmetry.

\par
Astrophysical black holes rotate. Angular momentum reduces the metric from static to stationary, and the cross term $g_{t\varphi}\neq0$ that appears in Boyer--Lindquist coordinates modifies each of the three conclusions above. The mechanism is superradiance. Zel'dovich pointed out that a rotating absorber can amplify an incident wave that satisfies a particular condition~\cite{Zeldovich:1971ffh}. Press and Teukolsky then recognized that the system would become a ``black-hole bomb'' if the amplified wave were repeatedly reflected back by some mechanism~\cite{Press:1972zz}. For a massive field this confinement is supplied by the mass term itself, which forms a barrier in the far region. A quasibound state therefore no longer decays but grows exponentially once $\omega<m\Omega_{H}$, where $\Omega_{H}=a/(r_{+}^{2}+a^{2})$ is the angular velocity of the horizon, and the cloud extracts energy and angular momentum from the rotation of the black hole. The instability was first quantified by Zouros and Eardley and by Detweiler by means of a matched asymptotic expansion, and was later computed systematically by Dolan with
Leaver's continued-fraction method~\cite{Zouros:1979iw,Detweiler:1980uk,Dolan:2007mj}. In the regime $\alpha\equiv GM_{\rm BH}M_{\rm boson}/\hbar c\lesssim1$, where $\alpha$ is the gravitational fine-structure constant, the bound-state spectrum takes a hydrogenic form, with Bohr radius $r_{B}=M/\alpha^{2}$ and levels $\omega\simeq\mu(1-\alpha^{2}/2n_{p}^{2})$; the system is accordingly known as a gravitational atom~\cite{Arvanitaki:2010sy,Arvanitaki:2014wva,Baumann:2018vus}. The consequences are observable. The cloud can grow to between $10^{-3}$ and $10^{-1}$ of the black-hole mass and radiates continuous gravitational waves at a frequency of about $2\omega_{R}$ through $\Phi^{2}$ annihilation, while the black hole spins down as it loses angular momentum and leaves a gap in the spin--mass plane; exclusion ranges for the ultralight boson mass have already been derived on this basis from measurements of black-hole spins~\cite{Baryakhtar:2017ngi,Ng:2020ruv}. The Kerr spacetime also admits exact results that have no counterpart in the spherically symmetric case: strictly stationary scalar clouds exist at the superradiance threshold $\omega=m\Omega_{H}$, and their self-gravitating extension yields Kerr black holes with scalar hair~\cite{Hod:2012px,Herdeiro:2014goa}. Rotation may therefore turn decay into growth, the lifetime problem into a growth-timescale problem, and a negligible energy into an observable cloud mass.

\par
The present work is motivated from two directions. On the one hand, studies of condensates retain the self-interaction $\lambda$ and thus deal with genuine condensates, but they are formulated exclusively on spherically symmetric static spacetimes, which structurally excludes superradiance. On the other hand, studies of superradiance and of the gravitational atom are formulated on the Kerr spacetime, but most of them set $\lambda=0$, so that their object is strictly a free massive scalar cloud rather than a condensate. The importance of the self-interaction has been recognized, for instance in the study of bosenova collapse by Yoshino and Kodama and in the analysis of level mixing and saturation by Baryakhtar \textit{et al.}; these discussions, however, mostly employ non-relativistic approximations or phenomenological estimates, and a Gross--Pitaevskii framework has not yet been set up and solved on a Kerr background starting from the relativistic Klein--Gordon equation~\cite{Baryakhtar:2017ngi,Yoshino:2012kn}. Three obstacles are encountered when a spherically symmetric background is generalized to an axisymmetric one. First, the Klein--Gordon equation remains separable in Kerr, but the angular equation is a spheroidal rather than a spherical harmonic equation, and its eigenvalue $A_{\ell m}$ depends on $c^{2}=a^{2}(\omega^{2}-\mu^{2})$, that is, on the frequency that is being sought; the radial equation in turn depends on the angular solution through $A_{\ell m}$, so that the eigenvalue problem is nonlinear and must be solved self-consistently by iteration. In the spherically symmetric case $A=\ell(\ell+1)$ is a constant and this coupling is absent~\cite{Carter:1968rr,Brill:1972xj,Teukolsky:1973ha}. Second, the Klein--Gordon equation in Kerr can be separated only after it has been multiplied throughout by $\Sigma=r^{2}+a^{2}\cos^{2}\theta$, whereupon the mass term $\mu^{2}\Sigma$ splits precisely into $\mu^{2}r^{2}$, which is assigned to
the radial part, and $\mu^{2}a^{2}\cos^{2}\theta$, which is assigned to the angular part. Once a self-interaction is included, $\mu^{2}$ is replaced by
$\mu^{2}+\lambda|\Phi|^{2}$; since $|\Phi|^{2}$ depends on both $r$ and $\theta$, it contaminates the two equations simultaneously and cannot be split. In the spherically symmetric case the solution is an $\ell=0$ state, $|\Phi|^{2}$ depends on $r$ alone, and the nonlinear term is merely an additional radial potential; this is precisely why the nonlinearity could be handled in earlier work. Third, Kerr possesses two Killing vectors and hence one additional conserved angular momentum. However, $\partial_{t}$ is spacelike inside the ergoregion, so that $-T^{t}{}_{t}$ is no longer a positive-definite local energy density, and the Klein--Gordon norm changes sign near the horizon in the superradiant regime; the interpretations of the energy and number densities given in the spherically symmetric literature must therefore be re-examined case by case.

\par
In this work a self-interacting Gross--Pitaevskii framework is set up and solved on a Kerr background, starting from the relativistic Klein--Gordon equation, in order to determine how the self-interaction modifies the structure, the shape and the superradiant evolution of an ultralight scalar cloud. The remainder of the paper is organized as follows. In Sec.~\ref{sec:2} the angular and radial equations and a closed-form effective potential are derived by separation of variables and a tortoise-coordinate transformation; the superradiance criterion is read off from the horizon limit of that potential, and two exact conservation relations are obtained together with the Gross--Pitaevskii equation of the single-mode projection. In Sec.~\ref{sec:3} the gravitational-atom spectrum of the quasibound states is obtained by a spectral method combined with a self-consistent eigenvalue solution in the cut-off potential, and the $\alpha^{4\ell+5}$ scaling law of the superradiant growth rate, together with the accuracy limits of that prescription, is computed from the conserved-current formula. In Sec.~\ref{sec:4} the Gross--Pitaevskii equation is solved by a bordered Newton method at fixed particle number, which yields the toroidal structure of the condensate, the domain of validity of the Thomas--Fermi approximation, and a criterion for superradiant saturation. Section~\ref{sec:5} summarizes the paper. Geometric units $G=c=\hbar=1$ are used throughout, lengths are measured in units of the ADM mass of the black hole, and the metric signature is taken to be $(-,+,+,+)$.

\section{Self-interacting scalar field in the Kerr spacetime}
\label{sec:2}

\subsection{Effective potential and the superradiance criterion}
\label{sec:2.1}
\par
In this section the radial equation obtained by separation of variables is brought into Schr\"odinger form, and the superradiance criterion is read off from the horizon limit of the effective potential. A rotating vacuum black hole of mass $M$ and angular momentum $J=aM$ is described by the Kerr solution, whose line element in Boyer--Lindquist (BL) coordinates $(t,r,\theta,\varphi)$ reads~\cite{Boyer:1966qh}
\begin{eqnarray}
ds^2=&-\Big(1-\frac{2Mr}{\Sigma}\Big)dt^2-\frac{4Mar\sin^2\theta}{\Sigma}\,dt\,d\varphi
+\frac{\Sigma}{\Delta}\,dr^2+\Sigma\,d\theta^2  \nonumber \\
&+\Big(r^2+a^2+\frac{2Ma^2r\sin^2\theta}{\Sigma}\Big)\sin^2\theta\,d\varphi^2 ,
\label{eq-2.1}
\end{eqnarray}
where the two basic functions are
\begin{equation}
\Delta(r)=r^2-2Mr+a^2, \qquad \Sigma(r,\theta)=r^2+a^2\cos^2\theta .
\label{eq-2.2}
\end{equation}
Equations (2.1) and (2.2) give the metric determinant $\sqrt{-g}=\Sigma\sin\theta$, whose proportionality to $\Sigma$ is what makes the separation of variables possible. Relative to a spherically symmetric static metric, the new structure in Eq.~(2.1) is the cross term $g_{t\varphi}=-2Mar\sin^{2}\theta/\Sigma$, that is, frame dragging. Inverting Eq.~(2.1), the components of the inverse metric are obtained as
\begin{eqnarray}
&g^{tt}=-\frac{(r^{2}+a^{2})^{2}-\Delta a^{2}\sin^{2}\theta}{\Sigma\Delta} ,
\qquad
g^{t\varphi}=-\frac{2Mar}{\Sigma\Delta} ,
\qquad
g^{\varphi\varphi}=\frac{\Delta-a^{2}\sin^{2}\theta}{\Sigma\Delta\sin^{2}\theta} , \nonumber \\
&g^{rr}=\frac{\Delta}{\Sigma} ,
\qquad
g^{\theta\theta}=\frac{1}{\Sigma} .
\label{eq-2.3}
\end{eqnarray}
As may be seen from Eq.~(2.3), $g^{tt}$, $g^{t\varphi}$ and $g^{\varphi\varphi}$ all contain $\Delta^{-1}$ and therefore diverge at the horizon; this coordinate singularity is dealt with below by means of the tortoise coordinate.

\par
\begin{figure*}[htbp]
\centering
\includegraphics[width=\textwidth]{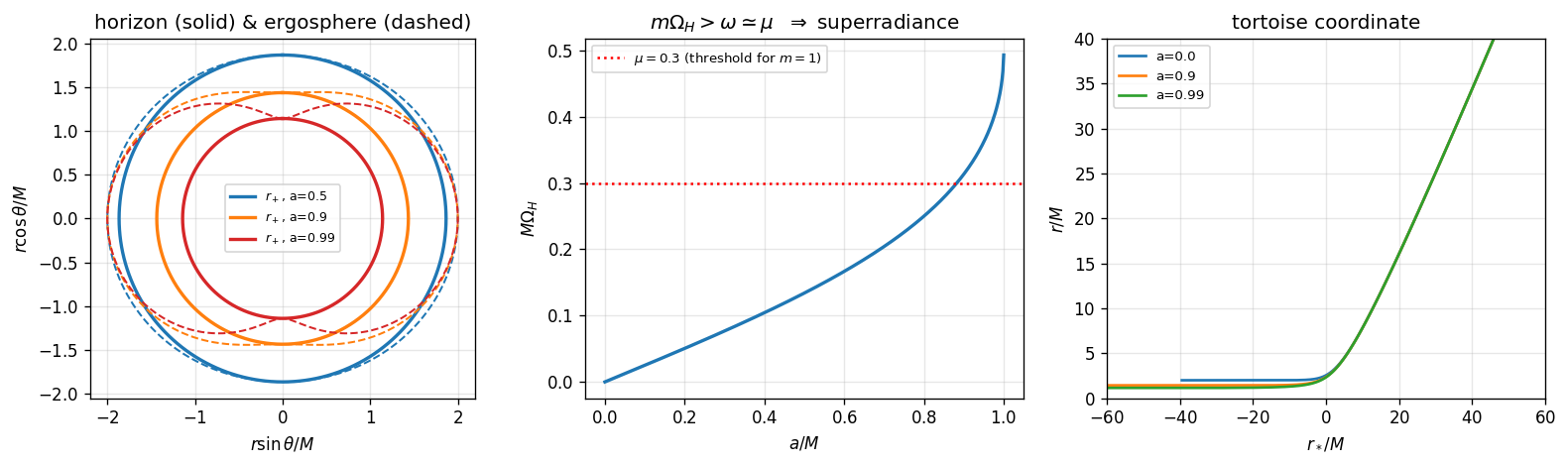}
\caption{Geometry of the Kerr spacetime. (a) Outer horizon (solid) and ergoregion boundary (dashed) in the meridional plane; the ergoregion widens towards the equator as the spin grows, while the two surfaces stay tangent on the axis. (b) Angular velocity of the horizon; superradiance with $m=1$ is possible only where $m\Omega_{H}$ lies above the dotted line. (c) Tortoise coordinate, which maps the horizon to $r_{*}\to-\infty$.}
\label{fig:1}
\end{figure*}

Setting $\Delta=0$ in Eq.~(2.2) locates the horizons and yields $r_{\pm}=M\pm\sqrt{M^{2}-a^{2}}$, $\Delta=(r-r_{+})(r-r_{-})$, together with the identity $r_{\pm}^{2}+a^{2}=2Mr_{\pm}$. Only the region outside the outer horizon, $r>r_{+}$, in which $\Delta>0$, is considered here. Setting $g_{tt}=0$ in Eq.~(2.1) instead locates the outer boundary of the ergoregion,
\begin{equation}
r_{\rm ergo}(\theta)=M+\sqrt{M^{2}-a^{2}\cos^{2}\theta} .
\label{eq-2.4}
\end{equation}
Inside the region enclosed by Eq.~(2.4) the Killing vector $\xi_{(t)}=\partial_{t}$ is spacelike, so that its conserved charge is no longer the energy measured by any observer~\cite{Wald:1984rg}. From $g_{t\varphi}$ and $g_{\varphi\varphi}$ of Eq.~(2.1) one obtains the angular velocity of the null generators of the horizon,
\begin{equation}
\Omega_{H}=-\frac{g_{t\varphi}}{g_{\varphi\varphi}}\bigg|_{r=r_{+}} =\frac{a}{r_{+}^{2}+a^{2}}=\frac{a}{2Mr_{+}} ,
\label{eq-2.5}
\end{equation}
where the identity above has been used in the second equality. The quantity $\Omega_{H}$ defined in Eq.~(2.5) is central to the superradiance criterion. In order to bring the radial equation into Schr\"odinger form, a coordinate is required that maps the physical region $r\in(r_{+},\infty)$ onto the whole real line; the tortoise coordinate is therefore defined by
\begin{equation}
\frac{dr_{*}}{dr}=\frac{r^{2}+a^{2}}{\Delta(r)} .
\label{eq-2.6}
\end{equation}
To integrate Eq.~(2.6), a partial-fraction decomposition is performed on its right-hand side and the residues are evaluated with the identity $r_{\pm}^{2}+a^{2}=2Mr_{\pm}$, which gives
\begin{equation}
\frac{r^{2}+a^{2}}{(r-r_{+})(r-r_{-})}
=1+\frac{1}{r_{+}-r_{-}}\Big[\frac{2Mr_{+}}{r-r_{+}}-\frac{2Mr_{-}}{r-r_{-}}\Big] .
\label{eq-2.7}
\end{equation}
Integrating Eq.~(2.7) term by term, the tortoise coordinate is obtained explicitly as
\begin{equation}
r_{*}=r+\frac{2Mr_{+}}{r_{+}-r_{-}}\ln\frac{r-r_{+}}{2M} -\frac{2Mr_{-}}{r_{+}-r_{-}}\ln\frac{r-r_{-}}{2M} .
\label{eq-2.8}
\end{equation}
The behaviour of Eq.~(2.8) at the two ends may now be examined. For $r\to\infty$ one has $r_{*}\simeq r+2M\ln r$, whereas for $r\to r_{+}$ the first logarithm dominates and inversion gives
\begin{equation}
r-r_{+}\simeq 2M\exp\!\Big[\frac{r_{+}-r_{-}}{2Mr_{+}}\,r_{*}\Big] .
\label{eq-2.9}
\end{equation}
For $a\to0$ Eq.~(2.8) reduces to the Schwarzschild expression $r_{*}=r+2M\ln(r/2M-1)$. Equation (2.9) shows that $r-r_{+}$ tends to zero exponentially in $r_{*}$: for $a=0.99M$ and $r_{*}=-200M$ one finds $r-r_{+}\sim10^{-11}M$, while at the same location $a=0$ gives $\sim e^{-101}$. The numerical inversion $r(r_{*})$ must therefore be carried out by Newton iteration on $\ln[(r-r_{+})/M]$ rather than on $r$ itself. The geometry described by Eqs.~(2.4), (2.5) and (2.8) is shown in Fig.~\ref{fig:1}.

\par
The matter field is introduced next. A complex scalar field with a self-interaction is considered, whose action reads~\cite{Chavanis:2011zi,Chavanis:2016shp}
\begin{equation}
S=-\int d^{4}x\sqrt{-g}\;\Big[g^{\mu\nu}\partial_{\mu}\Phi^{*}\partial_{\nu}\Phi+V(|\Phi|^{2})\Big] ,
\qquad
V=\mu^{2}|\Phi|^{2}+\frac{\lambda}{2}\big(|\Phi|^{2}\big)^{2} .
\label{eq-2.10}
\end{equation}
Varying Eq.~(2.10) with respect to $\Phi^{*}$ gives the field equation
\begin{equation}
\frac{1}{\sqrt{-g}}\,\partial_{\mu}\!\Big(\sqrt{-g}\,g^{\mu\nu}\partial_{\nu}\Phi\Big)
=\big(\mu^{2}+\lambda|\Phi|^{2}\big)\Phi .
\label{eq-2.11}
\end{equation}
A comparison of Eq.~(2.11) with the case $\lambda=0$ shows that the entire effect of the self-interaction is the replacement $\mu^{2}\to\mu^{2}+\lambda|\Phi|^{2}$. The dimensionless combination of the mass parameter $\mu=M_{\rm boson}c/\hbar$ in Eq.~(2.10) with the black-hole mass is the gravitational fine-structure constant,
\begin{equation}
\alpha\equiv\frac{GMM_{\rm boson}}{\hbar c}=M\mu
\;\xrightarrow{\;M=1\;}\;\alpha=\mu ,
\label{eq-2.12}
\end{equation}
which serves as the main expansion parameter of this work, while $\lambda$ is proportional to the scattering length. Two approximations are adopted here, in line with the treatment of the spherically symmetric case~\cite{Castellanos:2017tka}. The scalar field is not self-gravitating and the background is held fixed at the vacuum Kerr solution (2.1), which requires $E_{\rm cloud}/Mc^{2}\ll1$. The field is furthermore taken to be uncharged, so that it does not couple to an electromagnetic field; a charged field would require the minimal substitution $\partial_{\mu}\to\partial_{\mu}+i\hbar^{-1}qA_{\mu}$ in Eq.~(2.11).

\par
In the remainder of this subsection $\lambda=0$ is assumed. The two Killing vectors $\partial_{t}$ and $\partial_{\varphi}$ of Eq.~(2.1) allow the time and azimuthal dependences to be separated off, so that
\begin{equation}
\Phi=e^{-i\omega t+im\varphi}R(r)S(\theta) ,
\qquad m\in\mathbb{Z} ,
\label{eq-2.13}
\end{equation}
from which $\partial_{t}\Phi=-i\omega\Phi$ and $\partial_{\varphi}\Phi=im\Phi$ follow. Equations (2.3) and (2.13) are now inserted into Eq.~(2.11) and the result is multiplied throughout by $\Sigma$; the factor $\Sigma$ in $\sqrt{-g}=\Sigma\sin\theta$ is thereby cancelled against the $\Sigma^{-1}$ contained in $g^{rr}$ and $g^{\theta\theta}$, and one obtains
\begin{eqnarray}
&&\partial_{r}\big(\Delta\partial_{r}\Phi\big)
+\frac{1}{\sin\theta}\partial_{\theta}\big(\sin\theta\,\partial_{\theta}\Phi\big) \nonumber \\
&&\quad+\Big[\frac{\omega^{2}(r^{2}+a^{2})^{2}-4Mam\omega r+a^{2}m^{2}}{\Delta}
-\frac{m^{2}}{\sin^{2}\theta}-a^{2}\omega^{2}\sin^{2}\theta\Big]\Phi
=\mu^{2}\Sigma\,\Phi .
\label{eq-2.14}
\end{eqnarray}
The first two terms of Eq.~(2.14) already involve $r$ and $\theta$ separately, but the leading term in the square bracket and the right-hand side are not yet disentangled; two algebraic identities are required for their separation. To this end one introduces
\begin{equation}
\varpi(r)\equiv\omega(r^{2}+a^{2})-am ,
\label{eq-2.15}
\end{equation}
whose square is $\varpi^{2}=\omega^{2}(r^{2}+a^{2})^{2}-2am\omega(r^{2}+a^{2})+a^{2}m^{2}$. Since $-2am\omega(r^{2}+a^{2})+4Mam\omega r=-2am\omega(r^{2}+a^{2}-2Mr)=-2am\omega\Delta$, the two expressions combine and the leading bracket term of Eq.~(2.14) becomes
\begin{equation}
\frac{\omega^{2}(r^{2}+a^{2})^{2}-4Mam\omega r+a^{2}m^{2}}{\Delta}
=\frac{\varpi^{2}}{\Delta}+2am\omega .
\label{eq-2.16}
\end{equation}
The second term on the right of Eq.~(2.16) is independent of $\theta$ and free of $\Delta^{-1}$, so that it may be assigned directly to the radial part; the separation of the leading term is thereby completed. Next, following Eq.~(2.2), the right-hand side of Eq.~(2.14) is written as $\mu^{2}\Sigma=\mu^{2}r^{2}+\mu^{2}a^{2}\cos^{2}\theta$, its angular part is moved to the left and combined with $-a^{2}\omega^{2}\sin^{2}\theta$, and $\sin^{2}\theta=1-\cos^{2}\theta$ is used, which yields
\begin{equation}
-a^{2}\omega^{2}\sin^{2}\theta-\mu^{2}a^{2}\cos^{2}\theta
=-a^{2}\omega^{2}+a^{2}\big(\omega^{2}-\mu^{2}\big)\cos^{2}\theta .
\label{eq-2.17}
\end{equation}
Equation (2.17) splits the mass term between the two sides, $\mu^{2}r^{2}$ being assigned to the radial part and $\mu^{2}a^{2}\cos^{2}\theta$ to the angular one; it will be shown in Sec.~\ref{sec:2.2} that this step fails once $\lambda\neq0$, which is precisely where separability is lost. Substituting Eqs.~(2.16) and (2.17) back into Eq.~(2.14), the two sides involve $r$ and $\theta$ separately, and the introduction of a separation constant $A_{\ell m}$ yields two ordinary differential equations. The angular part gives the spheroidal harmonic equation
\begin{equation}
\frac{1}{\sin\theta}\frac{d}{d\theta}\Big(\sin\theta\frac{dS}{d\theta}\Big)
+\Big[c^{2}\cos^{2}\theta-\frac{m^{2}}{\sin^{2}\theta}+A_{\ell m}\Big]S=0 ,
\qquad
c^{2}\equiv a^{2}\big(\omega^{2}-\mu^{2}\big) ,
\label{eq-2.18}
\end{equation}
while the radial part gives the Teukolsky equation with $s=0$~\cite{Teukolsky:1973ha}
\begin{equation}
\Delta\frac{d}{dr}\Big(\Delta\frac{dR}{dr}\Big)
+\Big[\varpi^{2}-\Delta\big(\mu^{2}r^{2}+\lambda_{\ell m}\big)\Big]R=0 ,
\qquad
\lambda_{\ell m}\equiv A_{\ell m}+a^{2}\omega^{2}-2am\omega .
\label{eq-2.19}
\end{equation}
The separability of the scalar wave equation in Kerr originates in the existence of a Killing tensor.

\par
The properties of Eq.~(2.18) are discussed first. Its operator is self-adjoint for real $c^{2}$, so that $A_{\ell m}$ is real. For $c^{2}\to0$ the equation reduces to the associated Legendre equation, with $A_{\ell m}\to\ell(\ell+1)$ and $S\to P_{\ell}^{m}(\cos\theta)$. Treating $c^{2}$ as a small quantity, first order perturbation theory gives
\begin{equation}
A_{\ell m}=\ell(\ell+1)-c^{2}\big\langle\cos^{2}\theta\big\rangle_{\ell m}+O(c^{4}) ,
\qquad
\big\langle\cos^{2}\theta\big\rangle_{\ell m}=\frac{2\ell^{2}+2\ell-1-2m^{2}}{(2\ell-1)(2\ell+3)} .
\label{eq-2.20}
\end{equation}
Equation (2.20) shows that the departure of $A_{\ell m}$ from its spherical value is controlled by $c^{2}$. By the definition of $c^{2}$ in Eq.~(2.18), however, that quantity depends on the frequency $\omega$ that is being sought, while Eq.~(2.19) depends on $A_{\ell m}$ through $\lambda_{\ell m}$. The eigenvalue problem formed by the two equations is therefore nonlinear and must be solved by the self-consistent iteration
\begin{equation}
\omega\;\longrightarrow\;c^{2}=a^{2}(\omega^{2}-\mu^{2})\;\longrightarrow\;A_{\ell m}
\;\longrightarrow\;\omega
\label{eq-2.21}
\end{equation}
The iteration (2.21) degenerates for $a=0$, since $c^{2}\equiv0$ and $A_{\ell m}=\ell(\ell+1)$ is then a constant. This is the first technical difference between Kerr and a spherically symmetric background.

\par
Equation (2.19) is now brought into Schr\"odinger form so that the effective potential can be extracted. The transformation
\begin{equation}
\rho(r)\equiv\sqrt{r^{2}+a^{2}} ,
\qquad
R(r)=\frac{\psi(r)}{\rho(r)}
\label{eq-2.22}
\end{equation}
is introduced for this purpose. Equation (2.6) gives $\Delta\,d/dr=\rho^{2}\,d/dr_{*}$, so that the leading term of Eq.~(2.19) becomes $\rho^{2}\partial_{r_{*}}[\rho^{2}\partial_{r_{*}}(\psi/\rho)]$ once Eq.~(2.22) has been inserted. Differentiating successively, this expression is simplified by means of
\begin{equation}
\rho^{2}\,\partial_{r_{*}}\!\Big(\frac{\psi}{\rho}\Big)
=\rho\,\partial_{r_{*}}\psi-\psi\,\partial_{r_{*}}\rho ,
\qquad
\partial_{r_{*}}\big[\rho\,\partial_{r_{*}}\psi-\psi\,\partial_{r_{*}}\rho\big]
=\rho\,\partial_{r_{*}}^{2}\psi-\psi\,\partial_{r_{*}}^{2}\rho .
\label{eq-2.23}
\end{equation}
In the second of Eqs.~(2.23) the cross term $\partial_{r_{*}}\rho\,\partial_{r_{*}}\psi$ cancels exactly, which is precisely why $\rho=\sqrt{r^{2}+a^{2}}$ is chosen in Eq.~(2.22). Substituting Eq.~(2.23) back into Eq.~(2.19) and dividing by $\rho^{3}$, the standard form
\begin{equation}
\frac{d^{2}\psi}{dr_{*}^{2}}+\Big[\omega^{2}-V_{\rm eff}(r)\Big]\psi=0
\label{eq-2.24}
\end{equation}
is obtained, in which the effective potential is fixed by
\begin{equation}
\omega^{2}-V_{\rm eff}
=\frac{\varpi^{2}-\Delta\big(\mu^{2}r^{2}+\lambda_{\ell m}\big)}{(r^{2}+a^{2})^{2}}
-\frac{1}{\rho}\frac{d^{2}\rho}{dr_{*}^{2}} .
\label{eq-2.25}
\end{equation}
The last term of Eq.~(2.25) comes from $\psi\,\partial_{r_{*}}^{2}\rho$ in Eq.~(2.23) and has still to be evaluated explicitly. Equations (2.22) and (2.6) give $\partial_{r_{*}}\rho=\Delta r/\rho^{3}$; applying $\partial_{r_{*}}=(\Delta/\rho^{2})\partial_{r}$ once more and using $\Delta'=2(r-M)$, one finds
\begin{equation}
\frac{1}{\rho}\frac{d^{2}\rho}{dr_{*}^{2}}
=\frac{\Delta\big[2r(r-M)+\Delta\big]}{(r^{2}+a^{2})^{3}}-\frac{3\Delta^{2}r^{2}}{(r^{2}+a^{2})^{4}} .
\label{eq-2.26}
\end{equation}
Inserting Eq.~(2.26) into Eq.~(2.25) and solving for $V_{\rm eff}$, the effective potential is obtained in closed form,
\begin{equation}
V_{\rm eff}(r)=\omega^{2}
-\frac{\varpi^{2}-\Delta\big(\mu^{2}r^{2}+\lambda_{\ell m}\big)}{(r^{2}+a^{2})^{2}}
+\frac{\Delta\big[2r(r-M)+\Delta\big]}{(r^{2}+a^{2})^{3}}
-\frac{3\Delta^{2}r^{2}}{(r^{2}+a^{2})^{4}} .
\label{eq-2.27}
\end{equation}
It should be noted that Eq.~(2.27) depends explicitly on $\omega$ through
$\varpi$, $\lambda_{\ell m}$ and $A_{\ell m}(c^{2})$. Equation (2.24) is
therefore to be understood as the nonlinear eigenvalue problem
$-\psi''+V_{\rm eff}(r;\omega)\psi=\omega^{2}\psi$, in accordance with the
iterative structure of Eq.~(2.21). The spherically symmetric expression
$V_{\rm eff}=f(\mu^{2}+f'/r)$ carries no $\omega$
dependence~\cite{Castellanos:2017tka} and is the simplification obtained at
$a=0$.

\par
The physical content of Eq.~(2.27) resides in its two limits. For $r\to\infty$ one has $\varpi^{2}/(r^{2}+a^{2})^{2}\to\omega^{2}$ and $\Delta\mu^{2}r^{2}/(r^{2}+a^{2})^{2}\to\mu^{2}$, while the last two terms vanish as $O(r^{-2})$, so that
\begin{equation}
V_{\rm eff}(r\to\infty)=\mu^{2} .
\label{eq-2.28}
\end{equation}
Inserting Eq.~(2.28) into Eq.~(2.24) shows that an exponential solution at infinity requires $\omega^{2}<\mu^{2}$, which is the quasibound-state condition, the corresponding asymptotic behaviour being
\begin{equation}
\psi\sim e^{-\kappa r_{*}} ,
\qquad
\kappa\equiv\sqrt{\mu^{2}-\omega^{2}}
\qquad(r_{*}\to+\infty) .
\label{eq-2.29}
\end{equation}
A more accurate form carries a Coulomb-type correction, $\psi\sim r^{\nu}e^{-\kappa r}$ with $\nu=M(\mu^{2}-2\omega^{2})/\kappa$~\cite{Dolan:2007mj}. Inserting the condition $\omega^{2}<\mu^{2}$ into the definition of $c^{2}$ in Eq.~(2.18) gives $c^{2}<0$, whence Eq.~(2.20) yields $A_{\ell m}>\ell(\ell+1)$: the angular eigenvalue is always raised by the spin.

\par
At the other end, $\Delta\to0$ as $r\to r_{+}$, and the three terms of Eq.~(2.27) that contain $\Delta$ all vanish, leaving only the $\varpi^{2}$ term. To evaluate its limit, the identity $r_{+}^{2}+a^{2}=2Mr_{+}$ and Eq.~(2.5) are inserted into Eq.~(2.15), which gives
\begin{equation}
\varpi(r_{+})=\omega(r_{+}^{2}+a^{2})-am
=(r_{+}^{2}+a^{2})\big(\omega-m\Omega_{H}\big) .
\label{eq-2.30}
\end{equation}
Substituting Eq.~(2.30) into Eq.~(2.27) and letting $\Delta\to0$, the horizon limit of the effective potential is found to be
\begin{equation}
V_{\rm eff}(r_{+})=\omega^{2}-\big(\omega-m\Omega_{H}\big)^{2} .
\label{eq-2.31}
\end{equation}
Inserting Eq.~(2.31) into Eq.~(2.24) then gives the local wavenumber and the solution near the horizon,
\begin{equation}
k_{H}^{2}=\omega^{2}-V_{\rm eff}(r_{+})=\big(\omega-m\Omega_{H}\big)^{2} ,
\qquad
\psi\sim e^{-ik_{H}r_{*}}
\qquad(r_{*}\to-\infty) ,
\label{eq-2.32}
\end{equation}
where $k_{H}=\omega-m\Omega_{H}$. Of the two independent solutions in Eq.~(2.32), the physical boundary condition that nothing is emitted by the horizon requires the phase to be ingoing with respect to the co-rotating frame, which selects $e^{-ik_{H}r_{*}}$ uniquely and excludes $e^{+ik_{H}r_{*}}$. The direction of the number flux carried by this mode is nevertheless fixed by the sign of $k_{H}$. For $\omega>m\Omega_{H}$ one has $k_{H}>0$, the net flux is into the horizon, and the cloud decays; for $\omega<m\Omega_{H}$ one has $k_{H}<0$, the net flux is out of the horizon, and the cloud grows exponentially. The superradiance criterion is therefore $0<\omega<m\Omega_{H}$~\cite{Zeldovich:1971ffh,Press:1972zz}, and it follows directly from Eqs.~(2.30) and (2.31) without any numerical computation. For $a=0$, Eq.~(2.5) gives $\Omega_{H}=0$ and Eq.~(2.31) reduces to $V_{\rm eff}(r_{+})=0$; in that case $k_{H}=\omega>0$ always holds, the criterion admits no solution, and the cloud can only decay~\cite{Barranco:2012qs}.

\par
\begin{figure*}[htbp]
\centering
\includegraphics[width=\textwidth]{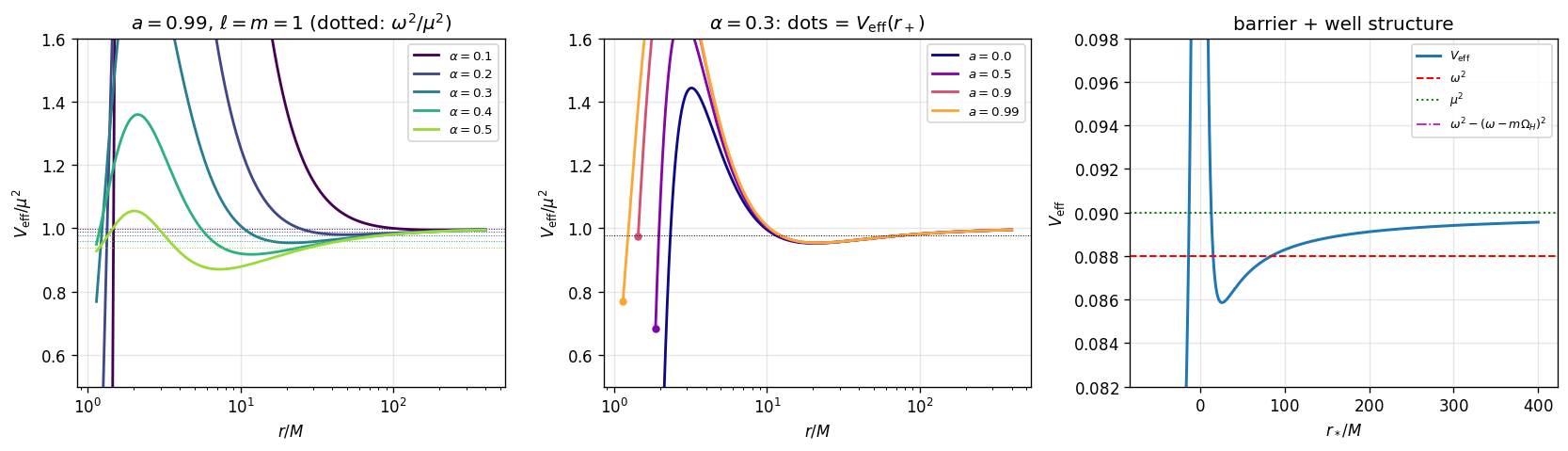}
\caption{Radial effective potential for $\ell=m=1$. (a) Dependence on $\alpha$ at $a=0.99M$; dotted lines of matching colour give the corresponding eigenfrequencies. A larger $\alpha$ deepens the well and lowers the barrier relative to $\mu^{2}$. (b) Dependence on the spin at fixed $\alpha=0.3$; the circles mark the horizon values. The spin modifies the potential only near the horizon. (c) The same potential against the tortoise coordinate, where the barrier and the well that traps the cloud are both apparent.}
\label{fig:2}
\end{figure*}

Equations (2.28) and (2.31) together determine the overall shape of $V_{\rm eff}$. Starting from the horizon value $\omega^{2}-(\omega-m\Omega_{H})^{2}$, the potential rises to a barrier peak $V_{\rm eff}(r_{\max})$ dominated by the centrifugal term $\Delta\lambda_{\ell m}/(r^{2}+a^{2})^{2}$ of Eq.~(2.27), falls to the bottom of a well $V_{\rm eff}(r_{\min})$ dominated by the gravitational attraction, and finally approaches $\mu^{2}$. For a quasibound state to be accommodated in that well, the eigenvalue of Eq.~(2.24) must satisfy
\begin{equation}
V_{\rm eff}(r_{\min})<\omega^{2}<\min\big\{\mu^{2},\;V_{\rm eff}(r_{\max})\big\} .
\label{eq-2.33}
\end{equation}
The barrier--well structure demanded by Eq.~(2.33), together with its dependence on $\alpha$ and $a$, is displayed in Fig.~\ref{fig:2}. Since the barrier is of finite height, Eq.~(2.24) admits no strictly square-integrable solution at real frequency; the frequency of an exact solution is $\omega=\omega_{R}+i\omega_{I}$, and the sign of $\omega_{I}$ is fixed by the superradiance criterion.

\par
\begin{figure*}[htbp]
\centering
\includegraphics[width=\textwidth]{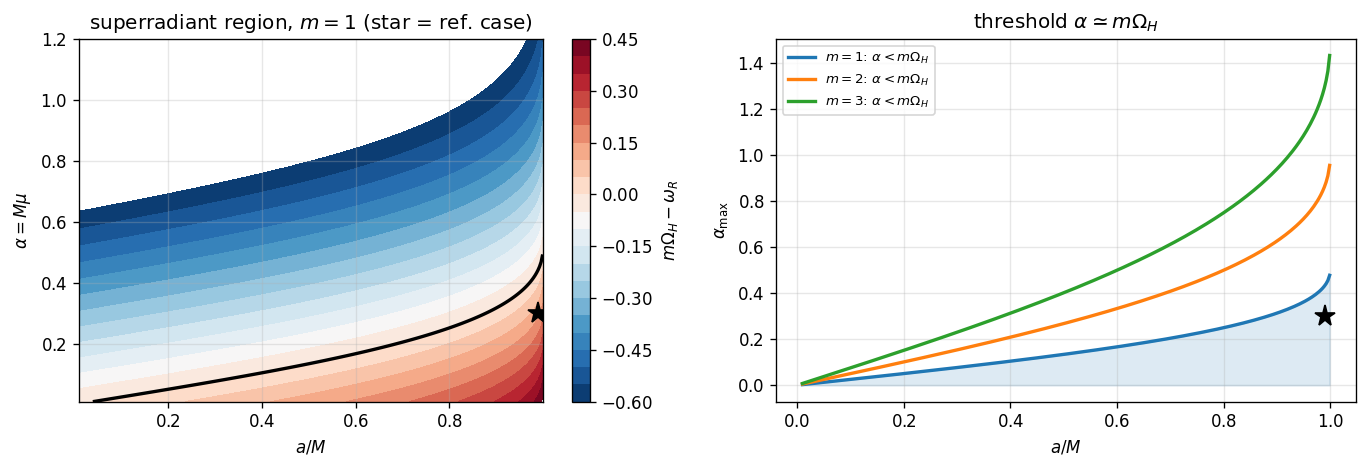}
\caption{Superradiance criterion in parameter space. (a) Unstable region for $m=1$: quasibound states grow in the red region and decay in the blue one, the black curve being the threshold and the star the reference configuration $a=0.99M$, $\alpha=0.3$. The threshold moves to larger $\alpha$ as the spin increases. (b) Threshold as a function of spin for $m=1,2,3$; the shaded band is the unstable region for $m=1$.}
\label{fig:3}
\end{figure*}

The realization of the superradiance criterion in parameter space is shown in Fig.~\ref{fig:3}. Since the real frequency of a quasibound state satisfies $\omega_{R}\simeq\mu$, the criterion reduces approximately to $\alpha\lesssim m\,M\Omega_{H}$, so that a large spin and a small $\alpha$ favour the instability. By Eq.~(2.5), $m\Omega_{H}$ increases monotonically with the spin and tends to $m/2M$ in the extremal Kerr limit. Although the threshold is raised linearly in $m$ for higher modes, their growth rate falls steeply with $\ell$, so that $\ell=m=1$ dominates in practice.

\par
The spherically symmetric limit of Eq.~(2.27) is checked last. Taking $a=0$ and $\ell=m=0$, Eq.~(2.19) gives $\lambda_{\ell m}=0$, while $\Delta=r^{2}f$, $\rho^{2}=r^{2}$ and $\varpi=\omega r^{2}$, with $f\equiv1-2M/r$. Inserting these into Eq.~(2.27) and collecting terms, one obtains
\begin{equation}
V_{\rm eff}=f\mu^{2}+\frac{f}{r^{2}}\Big[2-\frac{2M}{r}+f-3f\Big]
=f\mu^{2}+\frac{f}{r^{2}}\Big[2-\frac{2M}{r}-2f\Big] ,
\label{eq-2.34}
\end{equation}
and since the bracket in Eq.~(2.34) satisfies $2-2M/r-2f=2M/r$, it follows that
\begin{equation}
V_{\rm eff}\big|_{a=0,\,\ell=0}=f\mu^{2}+\frac{2Mf}{r^{3}}
=f(r)\Big[\mu^{2}+\frac{f'(r)}{r}\Big] .
\label{eq-2.35}
\end{equation}
Equation (2.35) agrees with the known result for the spherically symmetric case~\cite{Castellanos:2017tka}, and a numerical check gives a difference of no more than $3\times10^{-17}$ between the two, which validates the derivation of Eq.~(2.27).

\subsection{Conservation laws and the Gross--Pitaevskii equation}
\label{sec:2.2}
\par
The conservation laws implied by the symmetries of Eq.~(2.10) are established first, after which $\lambda\neq0$ is restored and separability is re-examined. Varying Eq.~(2.10) with respect to the metric gives the stress-energy tensor
\begin{equation}
T_{\mu\nu}=2\,\mathrm{Re}\big(\partial_{\mu}\Phi^{*}\partial_{\nu}\Phi\big)-g_{\mu\nu}\mathcal{L}_{0} ,
\qquad
\mathcal{L}_{0}\equiv g^{\rho\sigma}\partial_{\rho}\Phi^{*}\partial_{\sigma}\Phi+V ,
\label{eq-2.36}
\end{equation}
in which the scalar combination $\mathcal{L}_{0}$ has been introduced for brevity. Equation (2.10) is furthermore invariant under the global phase transformation $\Phi\to e^{i\chi}\Phi$, and the associated Noether current is
\begin{equation}
J^{\mu}=-i\big(\Phi^{*}\nabla^{\mu}\Phi-\Phi\nabla^{\mu}\Phi^{*}\big) ,
\qquad
\nabla_{\mu}J^{\mu}=0 .
\label{eq-2.37}
\end{equation}
The overall normalization of Eq.~(2.37) is fixed by the requirement that each particle carry an energy $\hbar\omega$, as may be verified with a plane wave $\Phi=Ae^{-i\omega t+ikx}$ in flat spacetime. The on-shell condition $\omega^{2}=k^{2}+\mu^{2}$ makes $\mathcal{L}_{0}=(-\omega^{2}+k^{2}+\mu^{2})|A|^{2}$ vanish in Eq.~(2.36), so that $-T^{t}{}_{t}=2\omega^{2}|A|^{2}$, while Eq.~(2.37) gives $J^{t}=2\omega|A|^{2}$; the ratio of the two is exactly $\omega$. This relation is shown below to hold in exact form in Kerr and for $\lambda\neq0$ as well. The two Killing vectors of Eq.~(2.1), together with the $U(1)$ symmetry of Eq.~(2.37), yield three conserved quantities, namely the particle number, the energy and the angular momentum, the last of which has no counterpart in the spherically symmetric case~\cite{Castellanos:2017tka}.

\par
The particle number is computed first. Inserting Eq.~(2.13) into Eq.~(2.37) and using $\Phi^{*}\partial_{t}\Phi-{\rm c.c.}=-2i\omega|\Phi|^{2}$ and $\Phi^{*}\partial_{\varphi}\Phi-{\rm c.c.}=2im|\Phi|^{2}$, together with $g^{tt}$ and $g^{t\varphi}$ of Eq.~(2.3), one obtains
\begin{equation}
J^{t}=2\omega\Big[-g^{tt}+\frac{m}{\omega}g^{t\varphi}\Big]|\Phi|^{2}
=\frac{2\omega}{\Sigma\Delta}
\Big[(r^{2}+a^{2})^{2}-\Delta a^{2}\sin^{2}\theta-\frac{2Mamr}{\omega}\Big]|\Phi|^{2} .
\label{eq-2.38}
\end{equation}
When Eq.~(2.38) is integrated, the factor $\Sigma^{-1}$ that it contains is cancelled against the $\Sigma$ of $\sqrt{-g}=\Sigma\sin\theta$. Using $|\Phi|^{2}=|S|^{2}|\psi|^{2}/(r^{2}+a^{2})$ from Eqs.~(2.13) and (2.22), and changing variables by $dr=\Delta(r^{2}+a^{2})^{-1}dr_{*}$ according to Eq.~(2.6), one finds
\begin{equation}
N=\int J^{t}\sqrt{-g}\,d^{3}x
=4\pi\omega\!\int_{-\infty}^{+\infty}\!|\psi|^{2}\,w(r)\,dr_{*} ,
\label{eq-2.39}
\end{equation}
where the weight function is given by
\begin{equation}
w(r)=1-\frac{\Delta\,a^{2}\big\langle\sin^{2}\theta\big\rangle+2Mamr/\omega}{(r^{2}+a^{2})^{2}} ,
\qquad
\big\langle\sin^{2}\theta\big\rangle\equiv\int_{0}^{\pi}|S|^{2}\sin^{3}\theta\,d\theta ,
\label{eq-2.40}
\end{equation}
the angular function being normalized as $\int_{0}^{\pi}|S|^{2}\sin\theta\,d\theta=1$. For $a\to0$ Eq.~(2.40) gives $w\to1$, and Eq.~(2.39) then reduces to the spherically symmetric expression $N=4\pi\!\int\!|u|^{2}f^{-1}dr$~\cite{Castellanos:2017tka}.

\par
The angular momentum is computed next. Equations (2.36) and (2.13) give $T_{t\varphi}=-2m\omega|\Phi|^{2}-g_{t\varphi}\mathcal{L}_{0}$ and $T_{\varphi\varphi}=2m^{2}|\Phi|^{2}-g_{\varphi\varphi}\mathcal{L}_{0}$, which upon insertion into $T^{t}{}_{\varphi}=g^{tt}T_{t\varphi}+g^{t\varphi}T_{\varphi\varphi}$ yield
\begin{equation}
T^{t}{}_{\varphi}
=2m\Big[-\omega g^{tt}+mg^{t\varphi}\Big]|\Phi|^{2}
-\big(g^{t\nu}g_{\nu\varphi}\big)\mathcal{L}_{0} .
\label{eq-2.41}
\end{equation}
The coefficient of the last term in Eq.~(2.41) is $g^{t\nu}g_{\nu\varphi}=\delta^{t}{}_{\varphi}=0$, so that $\mathcal{L}_{0}$ cancels completely. Comparison of what remains with Eq.~(2.38) gives $T^{t}{}_{\varphi}=mJ^{t}$, whose integral is
\begin{equation}
J_{z}=\int T^{t}{}_{\varphi}\sqrt{-g}\,d^{3}x=\hbar\,m\,N .
\label{eq-2.42}
\end{equation}
Equation (2.42) holds exactly and, because $\mathcal{L}_{0}$ has already cancelled, receives no correction from the self-interaction.

\par
The energy behaves differently. Equations (2.36) and (2.13) give $T_{tt}=2\omega^{2}|\Phi|^{2}-g_{tt}\mathcal{L}_{0}$ and $T_{\varphi t}=-2m\omega|\Phi|^{2}-g_{\varphi t}\mathcal{L}_{0}$; inserting these into $-T^{t}{}_{t}=-(g^{tt}T_{tt}+g^{t\varphi}T_{\varphi t})$ and comparing with Eq.~(2.38), one obtains
\begin{equation}
-T^{t}{}_{t}=-2\omega\Big[\omega g^{tt}-mg^{t\varphi}\Big]|\Phi|^{2}
+\big(g^{t\nu}g_{\nu t}\big)\mathcal{L}_{0}
=\omega J^{t}+\mathcal{L}_{0} .
\label{eq-2.43}
\end{equation}
In contrast with Eq.~(2.41), the coefficient of $\mathcal{L}_{0}$ in Eq.~(2.43) is $g^{t\nu}g_{\nu t}=\delta^{t}{}_{t}=1$, so that no cancellation occurs and the integral has to be evaluated separately. To this end Eq.~(2.11) is multiplied by $\Phi^{*}$ and integrated by parts, the boundary term being guaranteed to vanish by the exponential decay of Eq.~(2.29), which gives
\begin{equation}
\int g^{\mu\nu}\partial_{\mu}\Phi^{*}\partial_{\nu}\Phi\,\sqrt{-g}\,d^{3}x
=-\int\Big[\mu^{2}|\Phi|^{2}+\lambda|\Phi|^{4}\Big]\sqrt{-g}\,d^{3}x .
\label{eq-2.44}
\end{equation}
Substituting Eq.~(2.44) into the definition of $\mathcal{L}_{0}$ in Eq.~(2.36) and using $V$ of Eq.~(2.10), the quartic terms partially cancel and $\int\mathcal{L}_{0}\sqrt{-g}\,d^{3}x=-\tfrac{\lambda}{2}\int|\Phi|^{4}\sqrt{-g}\,d^{3}x$ is obtained. Integration of Eq.~(2.43) then gives
\begin{equation}
E=-\int T^{t}{}_{t}\sqrt{-g}\,d^{3}x
=\hbar\omega N-\frac{\lambda}{2}\int|\Phi|^{4}\sqrt{-g}\,d^{3}x .
\label{eq-2.45}
\end{equation}
Equations (2.42) and (2.45) show that each particle carries exactly $\hbar\omega$ and $\hbar m$, a single correction being contributed by the self-interaction to the energy alone. Its form matches the non-relativistic Bose--Einstein result $E=\mu N-\tfrac{g}{2}\!\int\!n^{2}$. Dividing the two relations gives the ratio of the angular momentum to the energy lost by the black hole as the cloud grows,
\begin{equation}
\frac{\Delta J}{\Delta E}=\frac{m}{\omega}\big[1+O(\lambda)\big] .
\label{eq-2.46}
\end{equation}
Since $\alpha\lesssim1$ by Eq.~(2.12), Eq.~(2.46) gives $m/\omega>m/\mu\gg1$. The relative loss of angular momentum is therefore much larger than that of mass, and the black hole mainly spins down. Once the spin has fallen to $m\Omega_{H}=\omega$, the superradiance criterion is no longer satisfied, the instability shuts off, and a gap is left in the spin--mass plane.

\par
Of the three conserved quantities, the number current also yields the growth rate. From $N\propto e^{2\omega_{I}t}$ one has $\omega_{I}=\dot N/2N$, so that the flux through the horizon has to be evaluated. Inserting $g^{rr}$ of Eq.~(2.3) and $\Delta\partial_{r}=(r^{2}+a^{2})\partial_{r_{*}}$ from Eq.~(2.6) into Eq.~(2.37), and noting that the derivative terms of $\sqrt{r^{2}+a^{2}}$ cancel in the antisymmetric combination, one obtains
\begin{equation}
\sqrt{-g}J^{r}=-i\sin\theta\,|S|^{2}
\big(\psi^{*}\partial_{r_{*}}\psi-\psi\partial_{r_{*}}\psi^{*}\big) .
\label{eq-2.47}
\end{equation}
At the horizon the asymptotic form $\psi\to A_{H}e^{-ik_{H}r_{*}}$ of Eq.~(2.32) is substituted into Eq.~(2.47), which gives $\psi^{*}\partial_{r_{*}}\psi-{\rm c.c.}=-2ik_{H}|A_{H}|^{2}$. Integrating over the sphere and dividing by $2N$ from Eq.~(2.39), one finds
\begin{equation}
\dot N=-4\pi k_{H}|A_{H}|^{2} ,
\qquad
\omega_{I}=-\frac{\big(\omega-m\Omega_{H}\big)\,|A_{H}|^{2}}
{2\,\omega\displaystyle\int|\psi|^{2}w(r)\,dr_{*} } .
\label{eq-2.48}
\end{equation}
Equation (2.48) realizes the superradiance criterion explicitly: $\omega_{I}$ carries the sign opposite to that of $k_{H}=\omega-m\Omega_{H}$, and for $a\to0$ the relation $k_{H}=\omega>0$ always gives $\omega_{I}<0$.

\par
Two sign issues that have no counterpart in the spherically symmetric case arise in Kerr from Eqs.~(2.36) and (2.38). First, $\partial_{t}$ is spacelike inside the ergoregion by Eq.~(2.4), and its conserved charge is not the energy measured by any observer; the quantity $-T^{t}{}_{t}$ of Eq.~(2.43) therefore remains a conserved Killing energy density, but it may take negative values. For a cloud localized at $r\sim M/\alpha^{2}\gg r_{\rm ergo}$, the contribution of the ergoregion is exponentially small. Second, evaluating the bracket of Eq.~(2.38) at the horizon with the identity $r_{+}^{2}+a^{2}=2Mr_{+}$ and Eq.~(2.5) gives
\begin{equation}
\Big[(r^{2}+a^{2})^{2}-\Delta a^{2}\sin^{2}\theta-\frac{2Mamr}{\omega}\Big]_{r\to r_{+}}
=\frac{2Mr_{+}(r_{+}^{2}+a^{2})}{\omega}\big(\omega-m\Omega_{H}\big) .
\label{eq-2.49}
\end{equation}
Equation (2.49) is negative wherever the superradiance criterion is met, so that the Klein--Gordon norm is not positive definite near the horizon. This is the formal origin of particle creation by superradiance, and it also shows that the weight $w(r)$ of Eq.~(2.40) may change sign there. For a localized cloud $|\psi|^{2}$ is exponentially small in that region, the correction for the reference case at $r\sim M/\alpha^{2}$ amounting to $O(10^{-8})$.

\par
The case $\lambda\neq0$ is now restored. By Eq.~(2.11), the only change relative to Sec.~\ref{sec:2.1} is that the mass term of Eq.~(2.17) becomes
\begin{equation}
\mu^{2}\Sigma\;\longrightarrow\;
\big(\mu^{2}+\lambda|\Phi|^{2}\big)r^{2}
+\big(\mu^{2}+\lambda|\Phi|^{2}\big)a^{2}\cos^{2}\theta .
\label{eq-2.50}
\end{equation}
It should be recalled from Eq.~(2.17) that the separation succeeds precisely because the two parts of $\mu^{2}\Sigma$ can be assigned separately to the radial and to the angular sector. Equation (2.13), however, gives $|\Phi|^{2}=|R(r)|^{2}|S(\theta)|^{2}$, which depends on $r$ and $\theta$ at the same time. The two terms of Eq.~(2.50) therefore become a $\theta$-dependent term $\lambda|R|^{2}|S|^{2}r^{2}$ in the radial equation and an $r$-dependent term $\lambda|R|^{2}|S|^{2}a^{2}\cos^{2}\theta$ in the angular one, and the two cannot be disentangled; the variables are thus not separable in Kerr once a self-interaction is present. In the spherically symmetric case $a=0$ and the solution is an $\ell=0$ state, so that $|S|^{2}$ is constant and $|\Phi|^{2}$ depends on $r$ alone. The nonlinear term of Eq.~(2.50) is then merely an additional radial potential and the one-dimensional structure is preserved, which is why the nonlinearity could be handled in earlier work; the property is lost as soon as angular momentum or rotation is introduced. Substituting Eq.~(2.50) into Eq.~(2.14), simplifying the leading term with Eq.~(2.16) and writing $\Phi=e^{-i\omega t+im\varphi}\Psi(r,\theta)$, one arrives at a two-dimensional nonlinear elliptic equation in which no approximation is involved,
\begin{eqnarray}
&&\partial_{r}\big(\Delta\partial_{r}\Psi\big)
+\frac{1}{\sin\theta}\partial_{\theta}\big(\sin\theta\,\partial_{\theta}\Psi\big)
+\Big[\frac{\varpi^{2}}{\Delta}+2am\omega
-\frac{m^{2}}{\sin^{2}\theta}-a^{2}\omega^{2}\sin^{2}\theta\Big]\Psi \nonumber \\
&&\qquad=\big(\mu^{2}+\lambda|\Psi|^{2}\big)\,\Sigma\,\Psi .
\label{eq-2.51}
\end{eqnarray}
Equation (2.51) is the basic equation of this work. Three strategies are available for solving it, and their relative merits are listed in Table~\ref{tab:routes}. Since the physics of interest here is the interplay between superradiance and the self-interaction, the first strategy is excluded and the second is adopted, the third being left for future work.
\begin{table*}[htbp]
\caption{Three strategies for solving the two-dimensional nonlinear equation. The present work adopts the second.}
\label{tab:routes}
\centering
\begin{tabular}{lll}
\hline\hline
strategy & advantage & cost \\
\hline
axisymmetric mode, $m=0$ & no mode coupling & $m\Omega_{H}=0$: superradiance excluded \\
single-mode projection & 1D ODE; superradiance retained & $\ell$-mixing neglected \\
full 2D numerics & most general & largest computational cost \\
\hline\hline
\end{tabular}
\end{table*}

\par
Following the second strategy, $\Psi=R(r)S_{\ell m}(\theta)$ is set, where $S_{\ell m}$ solves Eq.~(2.18) and is normalized as stated below Eq.~(2.40). Equation (2.51) is multiplied by $S_{\ell m}^{*}\sin\theta$ and integrated over $\theta$. The linear part gives $-A_{\ell m}$ by Eq.~(2.18), while the nonlinear term gives $\lambda|R|^{2}R\,\mathcal{I}(r)$, whose angular weight is fixed by the $\Sigma$ of Eq.~(2.2),
\begin{equation}
\mathcal{I}(r)\equiv\int_{0}^{\pi}|S|^{4}\Sigma\sin\theta\,d\theta
=r^{2}I_{4}+a^{2}I_{4}^{c} ,
\label{eq-2.52}
\end{equation}
the two angular moments appearing in Eq.~(2.52) being defined by
\begin{equation}
I_{4}\equiv\int_{0}^{\pi}|S|^{4}\sin\theta\,d\theta ,
\qquad
I_{4}^{c}\equiv\int_{0}^{\pi}|S|^{4}\cos^{2}\theta\,\sin\theta\,d\theta .
\label{eq-2.53}
\end{equation}
Collecting these results, the projected radial equation reads
\begin{equation}
\frac{d}{dr}\Big(\Delta\frac{dR}{dr}\Big)
+\Big[\frac{\varpi^{2}}{\Delta}+2am\omega-a^{2}\omega^{2}-A_{\ell m}-\mu^{2}r^{2}\Big]R
=\lambda\,\mathcal{I}(r)\,|R|^{2}R .
\label{eq-2.54}
\end{equation}
The steps of Eqs.~(2.22)--(2.24) are now repeated for Eq.~(2.54): the equation is multiplied by $\Delta$, $R=\psi/\rho$ is inserted, and a division by $\rho^{3}$ is carried out. The nonlinear term thereby becomes $\lambda\mathcal{I}\Delta|\psi|^{2}\psi/\rho^{6}$, and the Gross--Pitaevskii equation
\begin{equation}
-\frac{d^{2}\psi}{dr_{*}^{2}}+V_{\rm eff}(r;\omega)\,\psi
+\lambda_{\rm eff}(r)\,|\psi|^{2}\psi=\omega^{2}\psi ,
\qquad
\lambda_{\rm eff}(r)=\frac{\lambda\,\Delta(r)\,\mathcal{I}(r)}{(r^{2}+a^{2})^{3}}
\label{eq-2.55}
\end{equation}
is obtained, in which $V_{\rm eff}$ is still given by Eq.~(2.27). Equation (2.55) is the working equation of the sections that follow, and its $\lambda\to0$ limit is Eq.~(2.24).

\par
\begin{figure*}[htbp]
\centering
\includegraphics[width=\textwidth]{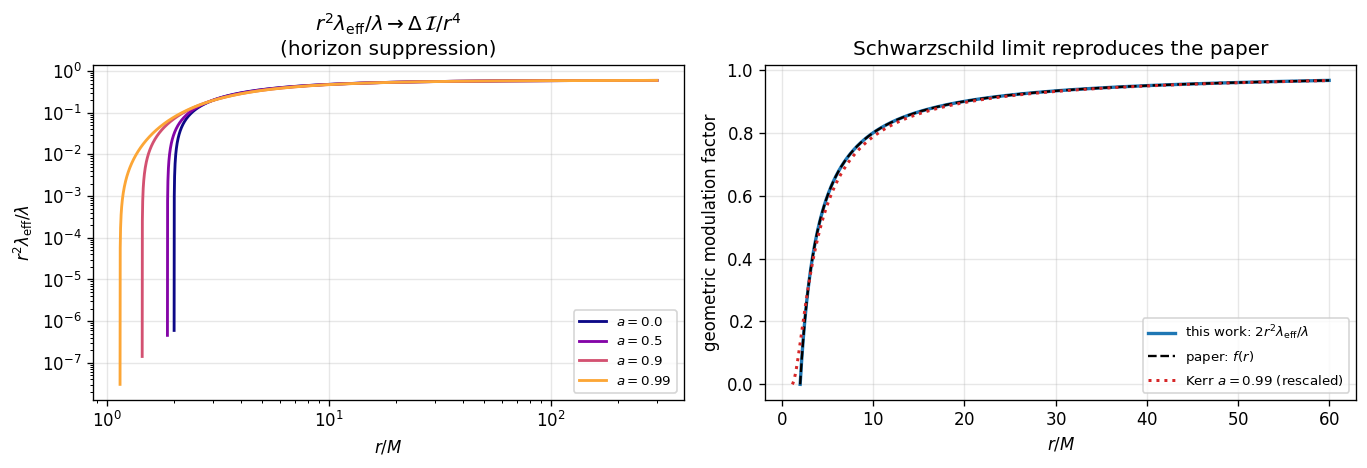}
\caption{Geometric modulation of the self-interaction strength. (a) Rescaled coupling $r^{2}\lambda_{\rm eff}/\lambda$ against $r/M$ for four spins at $\alpha=0.3$, $\ell=m=1$; the self-interaction is switched off at the horizon, and the spin shifts the radius at which this suppression sets in. (b) Check of the spherically symmetric limit: at $a=0$ and $\ell=m=0$ the rescaled coupling reproduces $f(r)=1-2M/r$, the $a=0.99M$ curve being shown for comparison.}
\label{fig:4}
\end{figure*}

The spherically symmetric limit of Eq.~(2.55) is checked last. For $a=0$ and $\ell=m=0$, Eq.~(2.18) gives a constant $S$, and the normalization requires $S=1/\sqrt2$; Eq.~(2.53) then gives $I_{4}=1/2$ and $I_{4}^{c}=1/6$, so that $\mathcal{I}=r^{2}/2$ follows from Eq.~(2.52). Using $\Delta=r^{2}f$ and $(r^{2}+a^{2})^{3}=r^{6}$, the effective self-interaction strength of Eq.~(2.55) reduces to
\begin{equation}
\lambda_{\rm eff}\big|_{a=0,\,\ell=0}=\frac{\lambda f(r)}{2r^{2}} .
\label{eq-2.56}
\end{equation}
The corresponding coefficient in the spherically symmetric case is $\lambda f/r^{2}$~\cite{Castellanos:2017tka}, which has the same structure as Eq.~(2.56); the factor of two arises from the normalization convention adopted for the angular function, since $\Phi=e^{i\omega t}u/r$ is used there, which is equivalent to $S\equiv1$. A numerical check gives a relative deviation of $\sim5\times10^{-8}$, limited by the accuracy of the angular quadrature. Since $\lambda_{\rm eff}\propto\Delta(r)$ in Eq.~(2.55), the self-interaction is suppressed by gravity and switched off at the horizon, the suppression factor $f(r)$ of the spherically symmetric case being replaced by $\Delta(r)/(r^{2}+a^{2})$ in Kerr. In addition, the term $a^{2}I_{4}^{c}$ in Eq.~(2.52) makes $\lambda_{\rm eff}$ depend on the angular structure of the cloud: $I_{4}$ and $I_{4}^{c}$ are determined by $S_{\ell m}$, which in turn depends on the spin and on the frequency through $c^{2}=a^{2}(\omega^{2}-\mu^{2})$ in Eq.~(2.18). A chain of dependence is thus established,
\begin{equation}
a\;\longrightarrow\;c^{2}\;\longrightarrow\;S_{\ell m}(\theta)
\;\longrightarrow\;I_{4},I_{4}^{c}\;\longrightarrow\;\lambda_{\rm eff}(r) .
\label{eq-2.57}
\end{equation}
The chain of Eq.~(2.57) degenerates in the spherically symmetric case, where $S$ is constant. The dependence of $\lambda_{\rm eff}$ on $r$ and $a$, together with the numerical check of its spherically symmetric limit, is shown in Fig.~\ref{fig:4}. Because $\lambda>0$ is repulsive, the self-interaction term of Eq.~(2.55) raises the chemical potential $\omega$, whereas the superradiance criterion requires $\omega<m\Omega_{H}$; a possible saturation mechanism follows from the two conditions taken together. The quasibound-state condition below Eq.~(2.28) requires $\omega<\mu$ at the same time, and it is shown in Sec.~\ref{sec:4.2} that the latter constraint usually operates first.

\section{Quasibound states and the superradiant instability}
\label{sec:3}
\subsection{Numerical spectrum and the gravitational-atom limit}
\label{sec:3.1}
\par
The eigenvalue problem posed by Eq.~(2.24) at $\lambda=0$ is solved in this section. Its solution serves both as the baseline against which the effects of the self-interaction are later judged, and as the place where the whole numerical implementation is tested. The reference case is taken to be
\begin{equation}
a=0.99M ,\qquad (\ell,m)=(1,1) ,\qquad \alpha=M\mu=0.30 ,\qquad n=0 ,
\label{eq-3.1}
\end{equation}
where $n$ denotes the number of radial nodes. Substituting Eq.~(3.1) into Eq.~(2.5) and into the horizon expressions given below Eq.~(2.7), one obtains
\begin{equation}
r_{+}=1.1411M ,\qquad M\Omega_{H}=0.4338 ,\qquad r_{B}\equiv M/\alpha^{2}=11.11M ,
\label{eq-3.2}
\end{equation}
in which $r_{B}$ is the gravitational Bohr radius~\cite{Arvanitaki:2009fg}. The mode $(\ell,m)=(1,1)$ has been chosen because it grows fastest under superradiance, while $a=0.99M$ and $\alpha=0.30$ balance the strength of the instability against the compactness of the cloud.

\par
The angular equation (2.18) is treated first. Writing $x=\cos\theta$ and defining the angular-momentum operator
\begin{equation}
\hat L^{2}\equiv-\frac{1}{\sin\theta}\partial_{\theta}\big(\sin\theta\,\partial_{\theta}\big)+\frac{m^{2}}{\sin^{2}\theta} ,
\label{eq-3.3}
\end{equation}
Eq.~(2.18) may be cast as the eigenvalue problem $(\hat L^{2}-c^{2}x^{2})S=A_{\ell m}S$. The function $S$ is expanded in the normalized associated Legendre basis,
\begin{equation}
S=\sum_{\ell'\ge|m|}b_{\ell'}\tilde P_{\ell'}^{\,m}(x) ,
\qquad
\tilde P_{\ell}^{\,m}\equiv\sqrt{\frac{2\ell+1}{2}\frac{(\ell-|m|)!}{(\ell+|m|)!}}\;P_{\ell}^{\,m} ,
\label{eq-3.4}
\end{equation}
which satisfies $\int_{0}^{\pi}\tilde P_{\ell}^{\,m}\tilde P_{\ell'}^{\,m}\sin\theta\,d\theta=\delta_{\ell\ell'}$ and $\hat L^{2}\tilde P_{\ell}^{\,m}=\ell(\ell+1)\tilde P_{\ell}^{\,m}$, so that the operator of Eq.~(3.3) is diagonal in this basis. The remaining term $c^{2}x^{2}$ is handled by the recurrence relation of the associated Legendre functions,
\begin{equation}
x\,\tilde P_{\ell}^{\,m}=c_{\ell}\,\tilde P_{\ell+1}^{\,m}+c_{\ell-1}\,\tilde P_{\ell-1}^{\,m} ,
\qquad
c_{\ell}=\sqrt{\frac{(\ell+1)^{2}-m^{2}}{(2\ell+1)(2\ell+3)}} ,
\label{eq-3.5}
\end{equation}
by which $x$ is represented as the tridiagonal matrix $X_{\ell\ell'}=c_{\ell}\delta_{\ell',\ell+1}+c_{\ell-1}\delta_{\ell',\ell-1}$. Inserting Eqs.~(3.4) and (3.5) into Eq.~(2.18), the angular problem is reduced to the real symmetric matrix eigenvalue problem
\begin{equation}
\mathbb{M}\,b=A_{\ell m}\,b ,
\qquad
\mathbb{M}=\mathrm{diag}\big[\ell(\ell+1)\big]-c^{2}X^{2} .
\label{eq-3.6}
\end{equation}
By Eq.~(3.5), $X^{2}$ connects $\ell$ only to $\ell$ and $\ell\pm2$, so that the matrix in Eq.~(3.6) is pentadiagonal and decouples according to the parity of $\ell-|m|$; its eigenvalues, taken in ascending order, correspond to $\ell=|m|,|m|+1,\dots$ The real symmetry of Eq.~(3.6) guarantees that $A_{\ell m}$ is real, in agreement with the statement made below Eq.~(2.18). The error made by truncating at $\ell\le\ell_{\max}$ decreases exponentially with $\ell_{\max}$, and a single diagonalization delivers the eigenvalues and eigenfunctions of all $\ell$ at once.

\par
The correctness of Eq.~(3.6) is confirmed by three checks. First, $\mathbb{M}$ is diagonal at $c^{2}=0$, and the numerical eigenvalues are $A=\{2,6,12,20,\dots\}$ for $m=1$, in agreement with the zeroth-order term of Eq.~(2.20). Second, the results at $c^{2}=10^{-5}$ are compared point by point with the first-order perturbative expression of Eq.~(2.20) in Table~\ref{tab:angular}; the deviations for four sets of quantum numbers all lie at the $10^{-13}$ level. Third, convergence in $\ell_{\max}$ is examined at the value $c^{2}=-4$ typical of a bound state, for $(\ell,m)=(1,1)$. The choices $\ell_{\max}=8,12,20,40,80$ give $A_{11}=2.73411102563441$, $2.73411102561226$, and three further values identical to the second, so that fifteen significant digits are stable once $\ell_{\max}\ge12$. All computations below are performed with $\ell_{\max}=30$--$40$. It may be noted that $A_{11}=2.734>\ell(\ell+1)=2$, which is consistent with the conclusion drawn from $c^{2}<0$ below Eq.~(2.29).
\begin{table*}[htbp]
\caption{Angular eigenvalue at $c^{2}=10^{-5}$, compared with first-order
perturbation theory.}
\label{tab:angular}
\centering
\begin{tabular}{cccc}
\hline\hline
$(\ell,m)$ & Eq.~(\ref{eq-3.6}) & Eq.~(\ref{eq-2.20}) & difference \\
\hline
$(1,1)$ & 1.999998000000 & 1.999998000000 & $4.6\times10^{-13}$ \\
$(2,1)$ & 5.999995714285 & 5.999995714286 & $3.8\times10^{-13}$ \\
$(2,2)$ & 5.999998571428 & 5.999998571429 & $2.0\times10^{-13}$ \\
$(3,0)$ & 11.999994888889 & 11.999994888889 & $3.2\times10^{-13}$ \\
\hline\hline
\end{tabular}
\end{table*}

\par
\begin{figure*}[htbp]
\centering
\includegraphics[width=\textwidth]{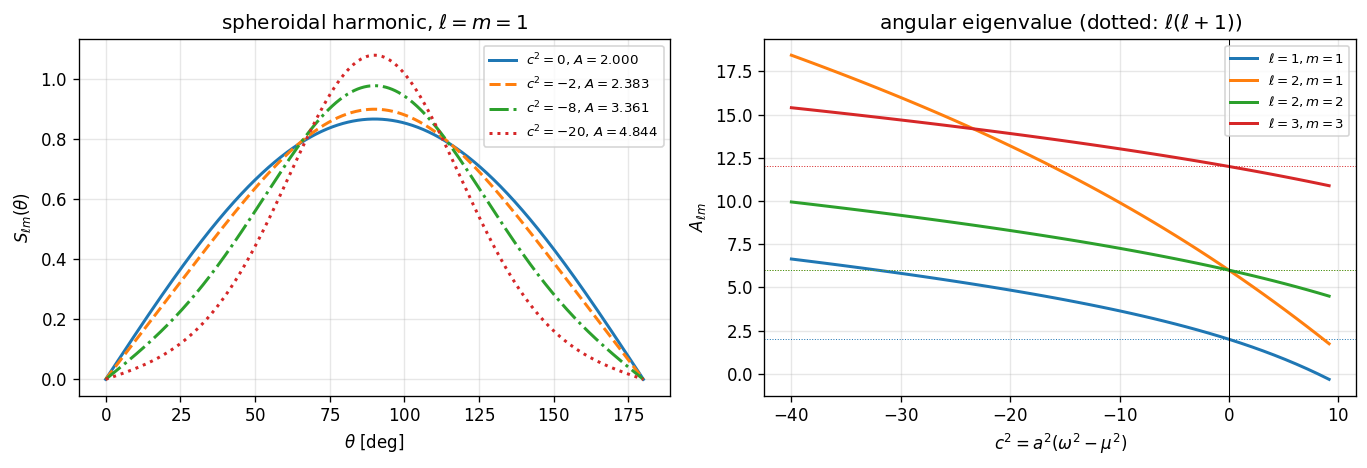}
\caption{Spheroidal harmonics obtained from the matrix eigenvalue problem. (a) Normalized angular function for $\ell=m=1$ at $c^{2}=0,-2,-8,-20$; as $c^{2}$ becomes more negative the function concentrates towards the equatorial plane. (b) Angular eigenvalue against $c^{2}$ for four sets of quantum numbers, the dotted lines marking the spherical values $\ell(\ell+1)$. Quasibound states lie on the $c^{2}<0$ side, where the eigenvalue is always raised.}
\label{fig:5}
\end{figure*}

The quantity $\langle\sin^{2}\theta\rangle$ of Eq.~(2.40) and the angular moments $I_{4}$ and $I_{4}^{c}$ of Eqs.~(2.52) and (2.53) are all evaluated from $S_{\ell m}$, and closed-form values are available in the limit $c^{2}\to0$ that may serve as a check. For $\ell=m=0$, Eq.~(2.18) gives $S=1/\sqrt2$, and direct integration yields $I_{4}=1/2$, $I_{4}^{c}=1/6$ and $\langle\sin^{2}\theta\rangle=2/3$. For $\ell=m=1$, Eq.~(2.18) gives $S\propto\sin\theta$, the normalization fixes $S=(\sqrt3/2)\sin\theta$, and substitution into Eq.~(2.53) yields
\begin{equation}
I_{4}=\frac{9}{16}\int_{0}^{\pi}\!\sin^{5}\theta\,d\theta=\frac{9}{16}\cdot\frac{16}{15}=\frac35 ,
\qquad
I_{4}^{c}=\frac{9}{16}\int_{0}^{\pi}\!\sin^{5}\theta\cos^{2}\theta\,d\theta=\frac{9}{16}\cdot\frac{16}{105}=\frac{3}{35} ,
\label{eq-3.7}
\end{equation}
while substitution into Eq.~(2.40) gives
\begin{equation}
\big\langle\sin^{2}\theta\big\rangle=\frac34\int_{0}^{\pi}\!\sin^{5}\theta\,d\theta=\frac45 .
\label{eq-3.8}
\end{equation}
Equations (3.7) and (3.8) are compared with the numerical results in Table~\ref{tab:moments}. Since $c^{2}=a^{2}(\omega^{2}-\mu^{2})=-2.16\times10^{-3}$ is small for the reference case (3.1), its angular moments differ from the closed-form values at $c^{2}\to0$ only at the $10^{-5}$ level, the corresponding eigenvalue shift being $A_{11}-\ell(\ell+1)=+4.3193\times10^{-4}$. The variation of the angular function and of its eigenvalue with $c^{2}$ is shown in Fig.~\ref{fig:5}.
\begin{table*}[htbp]
\caption{Angular moments computed from the numerical $S_{\ell m}$, compared with their closed-form values at $c^{2}\to0$. The $\ell=m=1$ column uses the reference case, for which $c^{2}=-2.16\times10^{-3}$.}
\label{tab:moments}
\centering
\begin{tabular}{ccccc}
\hline\hline
 & $\ell=m=0$ & exact & $\ell=m=1$ & exact \\
\hline
$I_{4}$ & 0.5000000257 & $1/2$ & 0.600030 & $3/5$ \\
$I_{4}^{c}$ & 0.1666666581 & $1/6$ & 0.085704 & $3/35$ \\
$\langle\sin^{2}\theta\rangle$ & 0.6666667009 & $2/3$ & 0.800020 & $4/5$ \\
\hline\hline
\end{tabular}
\end{table*}

\par
The radial problem is treated next. As discussed below Eq.~(2.33), the barrier is of finite height, so that Eq.~(2.24) admits no strictly square-integrable solution at real frequency. In order to reduce the problem to a self-adjoint one with real eigenvalues, the cut-off procedure of the spherically symmetric case is adopted and $V_{\rm eff}$ is levelled off inwards from the barrier peak $r_{\max}$,
\begin{equation}
\tilde V_{\rm eff}(r_{*})=
\begin{cases}
V_{\rm eff}\big(r_{\max}\big) , & r_{*}\le r_{*}^{\max} ,\\[3pt]
V_{\rm eff}\big(r(r_{*})\big) , & r_{*}>r_{*}^{\max} ,
\end{cases}
\label{eq-3.9}
\end{equation}
where $r_{*}^{\max}=r_{*}(r_{\max})$ is given by Eq.~(2.8). Within the constant-potential region defined by Eq.~(3.9) one has $\omega^{2}<V_{\rm eff}(r_{\max})$ by Eq.~(2.33), so that the solution decays exponentially,
\begin{equation}
\tilde\psi\propto e^{+\kappa_{\max}r_{*}} ,
\qquad
\kappa_{\max}=\sqrt{V_{\rm eff}(r_{\max})-\omega^{2}}
\qquad(r_{*}\to-\infty) ,
\label{eq-3.10}
\end{equation}
and, together with the decay at large distance given by Eq.~(2.29), the solution is rendered square integrable. The essential property of Eq.~(3.9) is that $\tilde V_{\rm eff}=V_{\rm eff}$ holds for $r_{*}>r_{*}^{\max}$, so that $\tilde\psi$ coincides with the true solution in that region and only the tunnelling information, that is $\omega_{I}$, is sacrificed. Equivalently, Eq.~(3.10) amounts to imposing a real logarithmic-derivative condition at $r_{*}^{\max}$,
\begin{equation}
\frac{\tilde\psi'}{\tilde\psi}\bigg|_{r_{*}^{\max}}=+\kappa_{\max} ,
\label{eq-3.11}
\end{equation}
whereas the logarithmic derivative of the true solution is complex there and is fixed by the one-way condition of Eq.~(2.32).

\par
A uniform grid $r_{*}^{(i)}=r_{*}^{\min}+ih$ is laid down on the tortoise coordinate, and the value $r^{(i)}=r(r_{*}^{(i)})$ at each node is obtained by the Newton iteration described below Eq.~(2.9). Applying second-order central differences to Eq.~(2.24) and replacing $V_{\rm eff}$ by $\tilde V_{\rm eff}$ of Eq.~(3.9), one obtains
\begin{equation}
-\frac{\psi_{i+1}-2\psi_{i}+\psi_{i-1}}{h^{2}}+\tilde V_{i}\psi_{i}=\omega^{2}\psi_{i} ,
\label{eq-3.12}
\end{equation}
subject to the Dirichlet conditions $\psi_{0}=\psi_{N-1}=0$. Equation (3.12) is the eigenvalue problem of the real symmetric tridiagonal matrix
\begin{equation}
\mathbb{T}=\mathrm{tridiag}\Big(-\frac{1}{h^{2}},\;\frac{2}{h^{2}}+\tilde V_{i},\;-\frac{1}{h^{2}}\Big) ,
\label{eq-3.13}
\end{equation}
from which only the $n$th eigenvalue is extracted by bisection combined with inverse iteration, at a cost of $O(N)$. The grid parameters are chosen according to the physical scales, the decay rate being
\begin{equation}
\kappa=\sqrt{\mu^{2}-\omega^{2}}\simeq\frac{\mu\alpha}{n_{p}}=\frac{\alpha^{2}}{n_{p}M} ,
\qquad n_{p}\equiv n+\ell+1 .
\label{eq-3.14}
\end{equation}
The outer boundary is therefore placed at $r_{*}^{\max\text{-grid}}=\Lambda/\kappa$, with $\Lambda=20$--$28$ in this work, that is, at twenty to twenty-eight decay lengths, while the inner boundary is set at $r_{*}^{\min}=-60M$ so as to accommodate the exponential decay of Eq.~(3.10). For the reference case (3.1), Eq.~(3.14) leads to the grid parameters
\begin{equation}
N=3\times10^{4} ,\qquad h=0.0205M ,\qquad r_{*}\in[-60M,\;556M] .
\label{eq-3.15}
\end{equation}

\par
\begin{figure}[htbp]
\centering
\includegraphics[width=9cm,height=6cm]{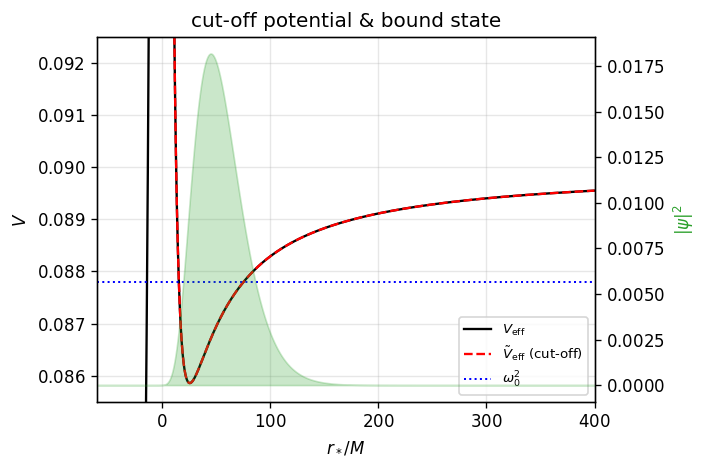}
\caption{Cut-off potential and the resulting quasibound state for the reference case. Left axis: the true effective potential (solid), the cut-off potential (dashed) and the eigenvalue. Right axis: the squared wave function (shaded). The two potentials coincide outside the barrier peak, which is where the cloud is localized, so that only the tunnelling is affected by the truncation.}
\label{fig:6}
\end{figure}

By Eq.~(2.21), $\tilde V_{\rm eff}$ depends on the frequency $\omega$ that is being sought through $A_{\ell m}(c^{2})$, so that Eq.~(3.12) has to be iterated jointly with the angular problem. The fixed-point iteration
\begin{equation}
\omega^{(k)}\;\xrightarrow{\;(2.18)\;}\;c^{2}\;\xrightarrow{\;(3.6)\;}\;A_{\ell m}\;
\xrightarrow{\;(2.19),(2.27),(3.9)\;}\;\tilde V_{\rm eff}\;\xrightarrow{\;(3.13)\;}\;\omega^{(k+1)}
\label{eq-3.16}
\end{equation}
is used for this purpose, the analytic spectrum of Eq.~(3.23) being taken as the starting value and $|\omega^{(k+1)}-\omega^{(k)}|<10^{-15}$ as the convergence criterion. One iteration is required at $\alpha=0.05$ and four at $\alpha=0.40$. This rapid convergence follows from the order of magnitude of $c^{2}$ in Eq.~(2.18),
\begin{equation}
|c^{2}|\simeq\frac{a^{2}\alpha^{4}}{n_{p}^{2}M^{2}}\ll1 ,
\label{eq-3.17}
\end{equation}
which gives $|c^{2}|=2.16\times10^{-3}$ for the reference case (3.1). Applying the grid of Eq.~(3.15) and the iteration of Eq.~(3.16) to the parameters (3.1), the ground-state eigenvalue is found to be
\begin{equation}
M\omega_{0}=0.296304586912 ,
\label{eq-3.18}
\end{equation}
with the corresponding binding energy and barrier parameters
\begin{equation}
\mu^{2}-\omega_{0}^{2}=2.2036\times10^{-3}M^{-2} ,
\qquad
r_{\max}=2.2515M ,
\qquad
V_{\rm eff}(r_{\max})=0.17587M^{-2} .
\label{eq-3.19}
\end{equation}
Equation (3.18) gives $\omega_{0}^{2}=0.087796M^{-2}$, which indeed satisfies the two-sided inequality of Eq.~(2.33), since $\mu^{2}=0.09M^{-2}$ while the barrier height in Eq.~(3.19) is $0.1759M^{-2}$. Equation (3.19) further shows that the barrier peak lies close to the horizon, $r_{\max}=2.25M$ against $r_{+}=1.14M$, whereas the wave function peaks at $r\approx39M$, that is, deep inside the well. The cut-off potential of Eq.~(3.9) and the resulting wave function are shown in Fig.~\ref{fig:6}.

\par
The value in Eq.~(3.18) may be compared with an analytic result. Generalizing Eq.~(2.35) to arbitrary $\ell$, corrections in $a$ being of higher order on a spherically symmetric background, and expanding in the far region, one obtains
\begin{equation}
V_{\rm eff}=f(r)\Big[\mu^{2}+\frac{\ell(\ell+1)}{r^{2}}+\frac{2M}{r^{3}}\Big]
=\mu^{2}-\frac{2M\mu^{2}}{r}+\frac{\ell(\ell+1)}{r^{2}}+O(r^{-3}) .
\label{eq-3.20}
\end{equation}
Inserting Eq.~(3.20) into Eq.~(2.24), setting $\omega=\mu(1+\epsilon)$ with $|\epsilon|\ll1$ so that $\omega^{2}-\mu^{2}\simeq2\mu^{2}\epsilon$, and noting that $r_{*}\simeq r$ in the far region, one finds
\begin{equation}
\frac{d^{2}\psi}{dr^{2}}+\Big[2\mu^{2}\epsilon+\frac{2M\mu^{2}}{r}-\frac{\ell(\ell+1)}{r^{2}}\Big]\psi=0 .
\label{eq-3.21}
\end{equation}
Equation (3.21) is the radial equation of the hydrogen atom. Comparing it term by term with the standard form $u''+[2/(a_{0}r)-1/(a_{0}^{2}n_{p}^{2})-\ell(\ell+1)/r^{2}]u=0$, the $1/r$ term fixes the Bohr radius and the constant term fixes the energy levels, so that
\begin{equation}
\frac{2}{a_{0}}=2M\mu^{2}\;\Longrightarrow\;r_{B}\equiv a_{0}=\frac{1}{M\mu^{2}}=\frac{M}{\alpha^{2}} ,
\qquad
2\mu^{2}\epsilon=-\frac{M^{2}\mu^{4}}{n_{p}^{2}}\;\Longrightarrow\;\epsilon=-\frac{\alpha^{2}}{2n_{p}^{2}} .
\label{eq-3.22}
\end{equation}
Substituting $\epsilon$ from Eq.~(3.22) back into $\omega=\mu(1+\epsilon)$, the gravitational-atom spectrum is obtained,
\begin{equation}
\omega_{n\ell m}=\mu\Big(1-\frac{\alpha^{2}}{2n_{p}^{2}}\Big)+O(\alpha^{5}\mu) ,
\qquad n_{p}=n+\ell+1 .
\label{eq-3.23}
\end{equation}
Equations (3.22) and (3.23) establish the formal correspondence between the present system and the hydrogen atom, which extends to the discrete spectrum, the principal quantum number and the Bohr radius; the fine structure and the selection rules at higher order are given in Ref.~\cite{Baumann:2018vus}. As a consistency check, the circular state of the hydrogen atom, $\ell=n_{p}-1$, satisfies $u\propto r^{n_{p}}e^{-r/n_{p}a_{0}}$ and its modulus squared peaks at $r=n_{p}^{2}a_{0}$. For the reference case, $n_{p}=2$ and $r_{B}=11.11M$ give $44.4M$, while the numerical solution of Eq.~(3.15) gives $39M$; the two differ by 12\%, which is comparable to the relativistic corrections of $O(\alpha^{2})=9\%$.

\par
The leading correction to Eq.~(3.23) is of relative order $\alpha^{4}$, and it arises both from the $O(r^{-3})$ term discarded in Eq.~(3.20) and from the next order of the expansion in $\epsilon$. The relative deviation of the numerical spectrum from Eq.~(3.23) should therefore obey
\begin{equation}
\delta\equiv\frac{|\omega_{\rm num}-\omega_{\rm hydro}|}{\omega_{\rm hydro}}=C_{4}\,\alpha^{4}+O(\alpha^{5}) ,
\label{eq-3.24}
\end{equation}
so that $\delta/\alpha^{4}$ should be approximately constant. Equation (3.24) is a test that involves the whole computational chain, since $\omega_{\rm num}$ depends at the same time on the effective potential of Eq.~(2.27), on the tortoise inversion described below Eq.~(2.9), on the cut-off potential of Eq.~(3.9), on the discretization of Eq.~(3.12), and on the self-consistent iteration of Eq.~(3.16). The results are listed in Table~\ref{tab:spectrum}, in which $\delta/\alpha^{4}$ is found to be stable within $\alpha\le0.2$ at
\begin{equation}
C_{4}(n=0,\ell=1)\simeq0.128 ,
\qquad
C_{4}(n=1,\ell=1)\simeq0.050 ,
\label{eq-3.25}
\end{equation}
whereas it rises by about 15\% at $\alpha=0.4$ owing to the $O(\alpha^{5})$ term of Eq.~(3.24). Table~\ref{tab:spectrum} also shows that the angular eigenvalue shift $A_{\ell m}-\ell(\ell+1)$ grows as $\alpha^{4}$ as well and remains positive throughout, in accordance with the estimate of Eq.~(3.17) and with the sign given by Eq.~(2.20); the number of self-consistent iterations increases slowly with $\alpha$, again as expected from Eq.~(3.17).
\begin{table*}[htbp]
\caption{Numerical spectrum compared with the hydrogenic formula, for
$a=0.99M$ and $\ell=m=1$. The column $\delta/\alpha^{4}$ tests the predicted
scaling.}
\label{tab:spectrum}
\centering
\begin{tabular}{cccccccc}
\hline\hline
$\alpha$ & $n$ & $M\omega_{\rm num}$ & $M\omega_{\rm hydro}$ & $\delta$ &
$\delta/\alpha^{4}$ & $A_{\ell m}-\ell(\ell+1)$ & iterations \\
\hline
0.05 & 0 & 0.049984334546 & 0.0499843750 & $8.09\times10^{-7}$ & 0.1295 & $+3.07\times10^{-7}$ & 1 \\
0.05 & 1 & 0.049993039933 & 0.0499930556 & $3.13\times10^{-7}$ & 0.0500 & $+1.36\times10^{-7}$ & 1 \\
0.10 & 0 & 0.099873725357 & 0.0998750000 & $1.28\times10^{-5}$ & 0.1276 & $+4.95\times10^{-6}$ & 2 \\
0.10 & 1 & 0.099943948710 & 0.0999444444 & $4.96\times10^{-6}$ & 0.0496 & $+2.20\times10^{-6}$ & 2 \\
0.20 & 0 & 0.198959317234 & 0.1990000000 & $2.04\times10^{-4}$ & 0.1278 & $+8.14\times10^{-5}$ & 3 \\
0.20 & 1 & 0.199539487328 & 0.1995555556 & $8.05\times10^{-5}$ & 0.0503 & $+3.61\times10^{-5}$ & 3 \\
0.30 & 0 & 0.296304586962 & 0.2966250000 & $1.08\times10^{-3}$ & 0.1334 & $+4.32\times10^{-4}$ & 4 \\
0.30 & 1 & 0.298370995624 & 0.2985000000 & $4.32\times10^{-4}$ & 0.0534 & $+1.91\times10^{-4}$ & 3 \\
0.40 & 0 & 0.390529286391 & 0.3920000000 & $3.75\times10^{-3}$ & 0.1466 & $+1.47\times10^{-3}$ & 4 \\
0.40 & 1 & 0.395834802994 & 0.3964444444 & $1.54\times10^{-3}$ & 0.0601 & $+6.50\times10^{-4}$ & 4 \\
\hline\hline
\end{tabular}
\end{table*}

\par
\begin{figure*}[htbp]
\centering
\includegraphics[width=\textwidth]{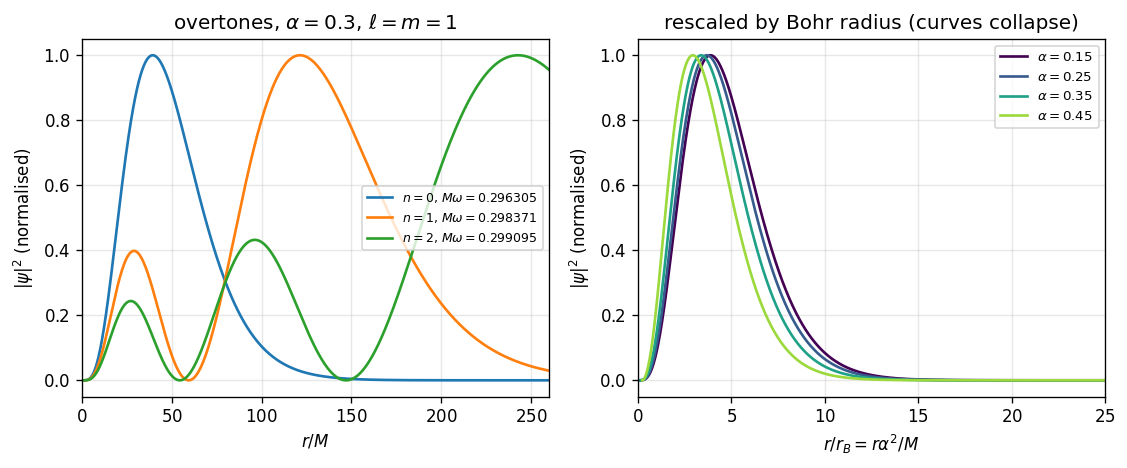}
\caption{Quasibound-state wave functions. (a) Normalized $|\psi|^{2}$ against $r/M$ for $\alpha=0.3$, $\ell=m=1$ and overtone numbers $n=0,1,2$; the state carries $n$ radial nodes and its extent grows with the principal quantum number. (b) The same against the radius rescaled by the gravitational Bohr radius, for $\alpha=0.15$--$0.45$ at $n=0$. The near collapse shows that the size of the cloud is set by that radius alone, the residual spread measuring the relativistic corrections.}
\label{fig:7}
\end{figure*}

Equation (3.22) further shows that the size of the cloud is set entirely by $r_{B}=M/\alpha^{2}$. Wave functions obtained at different $\alpha$ should therefore collapse onto a single curve in the limit $\alpha\to0$ when plotted against $r/r_{B}$, and this is confirmed numerically, the residual spread being due to relativistic corrections. The role of the overtone number $n$ in Eq.~(3.23) is likewise the same as in the hydrogen atom: the wave function carries $n$ radial nodes, its extent grows as $n_{p}^{2}$, and the binding energy falls as $n_{p}^{-2}$. Both properties are displayed in Fig.~\ref{fig:7}.

\subsection{Superradiant growth rate and accuracy calibration}
\label{sec:3.2}
\par
The growth rate has already been given by Eq.~(2.48), in which the only unknown is the horizon amplitude $|A_{H}|$. By the construction of Sec.~\ref{sec:3.1}, the cut-off solution of Eq.~(3.9) coincides with the true solution for $r_{*}>r_{*}^{\max}$. Its value and derivative at $r_{*}^{\max}$, $(\tilde\psi,\tilde\psi')$, may therefore be used as initial data for an inward integration performed with the untruncated $V_{\rm eff}$,
\begin{equation}
\frac{d^{2}\psi}{dr_{*}^{2}}=\big[V_{\rm eff}(r)-\omega^{2}\big]\psi ,
\qquad r_{*}:\;r_{*}^{\max}\to r_{*}^{\rm deep}=-450M .
\label{eq-3.26}
\end{equation}
Inside the barrier the growing solution dominates on inward integration, so that this direction is numerically stable for Eq.~(3.26). Once the constant-potential region near the horizon has been reached, the two independent solutions of Eq.~(2.32) combine into a real solution whose envelope is read off from
\begin{equation}
\psi\longrightarrow\mathcal{A}\cos\big(k_{H}r_{*}+\delta\big) ,
\qquad
\mathcal{A}=\sqrt{\psi^{2}+\big(\psi'/k_{H}\big)^{2}} ,
\label{eq-3.27}
\end{equation}
where $\mathcal{A}$ is exactly conserved within that region and may thus be evaluated at any depth. The standing wave of Eq.~(3.27) is a superposition of ingoing and outgoing components of equal amplitude, the ingoing one having amplitude $|A_{H}|=\mathcal{A}/2$. Substituting this into Eq.~(2.48) gives
\begin{equation}
\omega_{I}=-\frac{\big(\omega-m\Omega_{H}\big)\,\mathcal{A}^{2}}
{8\,\omega\displaystyle\int_{r_{*}^{\max}}^{\infty}|\psi|^{2}w(r)\,dr_{*}} ,
\label{eq-3.28}
\end{equation}
with $w(r)$ given by Eq.~(2.40). The relative standard deviation of the envelope (3.27) over $r_{*}\in[-450M,-300M]$ ranges from $7\times10^{-12}$ at $\alpha=0.30$ to $1.8\times10^{-11}$ at $\alpha=0.08$, which shows that the constant-potential region is well established and that the integration of Eq.~(3.26) is sufficiently accurate.

\par
An independent comparison is required for Eq.~(3.28), and the matched asymptotic result of Detweiler is adopted for that purpose. Its general form is~\cite{Detweiler:1980uk}
\begin{equation}
M\omega_{I}=2\tilde r_{+}\,C_{n\ell m}\,\big(m\Omega_{H}-\omega_{R}\big)\,\alpha^{4\ell+5} ,
\label{eq-3.29}
\end{equation}
where $\tilde a=a/M$ and $\tilde r_{+}=r_{+}/M$, $\omega_{R}$ is taken from the spectrum of Eq.~(3.23), and the coefficient is
\begin{equation}
C_{n\ell m}=\frac{2^{4\ell+2}(2\ell+n+1)!}{(n+\ell+1)^{2\ell+4}\,n!}
\left[\frac{\ell!}{(2\ell)!(2\ell+1)!}\right]^{2}
\prod_{k=1}^{\ell}\Big[k^{2}\big(1-\tilde a^{2}\big)+\big(\tilde am-2\tilde r_{+}\alpha\big)^{2}\Big] .
\label{eq-3.30}
\end{equation}
Setting $\ell=m=1$ and $n=0$ in Eq.~(3.30), the factorials give $1/24$ and only one factor survives in the product, so that
\begin{equation}
C_{011}=\frac{1}{24}\Big[\big(1-\tilde a^{2}\big)+\big(\tilde a-2\tilde r_{+}\alpha\big)^{2}\Big] .
\label{eq-3.31}
\end{equation}
Inserting Eq.~(3.23) into the bracket of Eq.~(3.29) gives $2\tilde r_{+}(m\Omega_{H}-\omega_{R})M=\tilde a-2\tilde r_{+}\alpha+O(\alpha^{3})$, so that Eqs.~(3.29) and (3.31) reduce for $\tilde a\ll1$ to Detweiler's original expression $\omega_{I}\simeq(\tilde a-2\mu r_{+})\mu\alpha^{8}/24$ \cite{Detweiler:1980uk}. It should be emphasized that Eq.~(3.29) is valid only for $\alpha\ll1$; near the peak of the growth rate, at $\alpha\sim0.4$, it may depart from the exact result by an order of magnitude.

\par
\begin{figure*}[htbp]
\centering
\includegraphics[width=\textwidth]{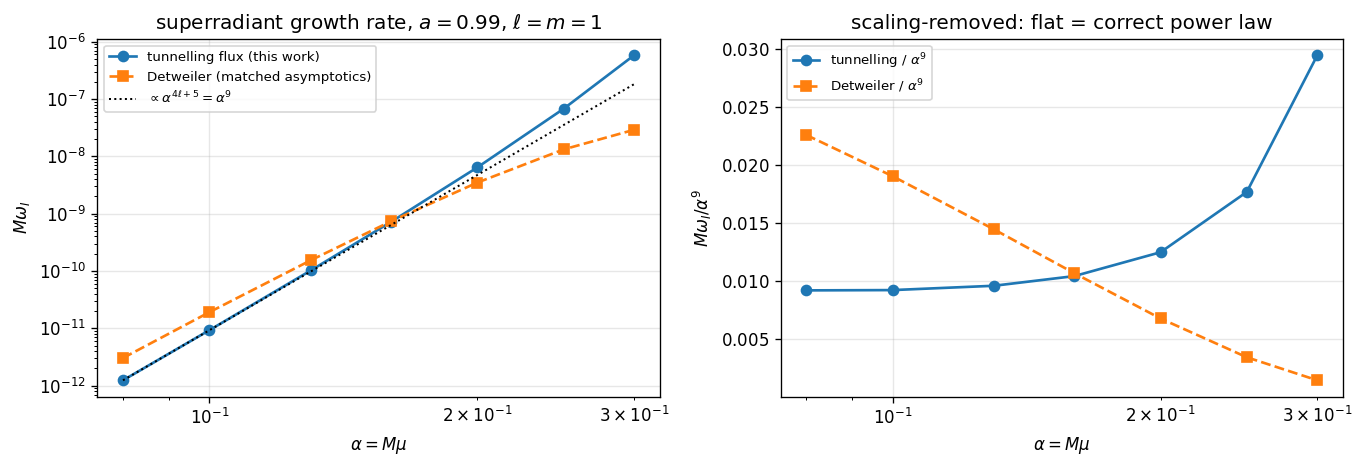}
\caption{Superradiant growth rate for the fastest-growing mode. (a) $M\omega_{I}$ against $\alpha$ on logarithmic axes: circles give the flux formula of this work, squares the matched asymptotic expression, and the dotted line a reference slope $\alpha^{9}$. (b) The same data with that scaling divided out. The flux result flattens for $\alpha\lesssim0.13$, which confirms the power law; the vertical offset between the two curves is the $O(1)$ coefficient discrepancy discussed in the text.}
\label{fig:8}
\end{figure*}

The point-by-point comparison of Eqs.~(3.28) and (3.29) is presented in Table~\ref{tab:growth}, all the values of $\alpha$ listed there satisfying the superradiance criterion given below Eq.~(2.31). Once the growth rate has been divided by $\alpha^{9}$, a constant is approached at small $\alpha$,
\begin{equation}
\frac{M\omega_{I}}{\alpha^{9}}=9.22\times10^{-3},\;9.25\times10^{-3},\;9.62\times10^{-3}
\qquad(\alpha=0.08,\,0.10,\,0.13) ,
\label{eq-3.32}
\end{equation}
the three values agreeing to within a few per cent. The analytic scaling law
\begin{equation}
\omega_{I}\propto\alpha^{4\ell+5}\;\xrightarrow{\;\ell=1\;}\;\alpha^{9}
\label{eq-3.33}
\end{equation}
is thereby reproduced independently by Eq.~(3.28), while $\omega_{I}$ itself spans almost three orders of magnitude over that range. Since Eqs.~(3.28) and (3.29) are derived independently, the former from the conserved current of Eq.~(2.37) and the numerical integration of Eq.~(3.26), the latter from a matched asymptotic expansion, Eq.~(3.32) tests the entire chain formed by the
effective potential of Eq.~(2.27), the boundary condition of Eq.~(2.32) and the conserved currents of Eqs.~(2.38)--(2.48). The scaling law (3.33) and the comparison of the two methods are shown in Fig.~\ref{fig:8}.
\begin{table*}[htbp]
\caption{Superradiant growth rate for $a=0.99M$, $\ell=m=1$ and $n=0$, all
listed values of $\alpha$ being superradiant. The last column is the ratio of
the two methods.}
\label{tab:growth}
\centering
\begin{tabular}{cccccc}
\hline\hline
 & & & \multicolumn{2}{c}{$M\omega_{I}$} & \\
\cline{4-5}
$\alpha$ & $M\omega_{R}$ & $|A_{H}|$ &
Eq.~(\ref{eq-3.28}) & Eq.~(\ref{eq-3.29}) & ratio \\
\hline
0.08 & 0.079936 & $7.48\times10^{-7}$ & $1.238\times10^{-12}$ & $3.034\times10^{-12}$ & 0.408 \\
0.10 & 0.099874 & $2.35\times10^{-6}$ & $9.246\times10^{-12}$ & $1.906\times10^{-11}$ & 0.485 \\
0.13 & 0.129721 & $9.33\times10^{-6}$ & $1.020\times10^{-10}$ & $1.535\times10^{-10}$ & 0.665 \\
0.16 & 0.159475 & $2.89\times10^{-5}$ & $7.181\times10^{-10}$ & $7.356\times10^{-10}$ & 0.976 \\
0.20 & 0.198959 & $1.04\times10^{-4}$ & $6.403\times10^{-9}$  & $3.482\times10^{-9}$  & 1.839 \\
0.25 & 0.247921 & $4.24\times10^{-4}$ & $6.752\times10^{-8}$  & $1.320\times10^{-8}$  & 5.116 \\
0.30 & 0.296305 & $1.58\times10^{-3}$ & $5.801\times10^{-7}$  & $2.905\times10^{-8}$  & 19.968 \\
\hline\hline
\end{tabular}
\end{table*}

\par
The sign of Eq.~(3.28) must also be checked. As analysed below Eq.~(2.32), $\omega_{I}<0$ is expected whenever $\omega>m\Omega_{H}$. In order to realize that situation, $\alpha=0.20$ and $\ell=m=1$ are held fixed while the spin is lowered, so that $m\Omega_{H}$ of Eq.~(2.5) falls below $\omega$; the results are collected in Table~\ref{tab:sign}. The three spins all give a negative $\omega_{I}$ whose sign is strictly opposite to that of $k_{H}$, which confirms the criterion implied by Eqs.~(2.31) and (3.28). It may be noted that the two methods agree considerably better in this non-superradiant regime, the difference being 11\% at $a=0.2M$; no near cancellation of $(m\Omega_{H}-\omega_{R})$ occurs in Eq.~(3.29) here, and the discrepancy is then governed by the thickness of the barrier alone.
\begin{table*}[htbp]
\caption{Sign check in the non-superradiant regime, at fixed $\alpha=0.20$ and $\ell=m=1$ with decreasing spin.}
\label{tab:sign}
\centering
\begin{tabular}{cccc}
\hline\hline
$a/M$ & $m\Omega_{H}$ & $M\omega_{I}$, Eq.~(\ref{eq-3.28}) &
$M\omega_{I}$, Eq.~(\ref{eq-3.29}) \\
\hline
0.10 & 0.02506 & $-1.881\times10^{-8}$ & $-2.187\times10^{-8}$ \\
0.20 & 0.05051 & $-1.824\times10^{-8}$ & $-1.644\times10^{-8}$ \\
0.30 & 0.07677 & $-1.843\times10^{-8}$ & $-1.164\times10^{-8}$ \\
\hline\hline
\end{tabular}
\end{table*}

\par
Although the scaling laws of Eqs.~(3.32) and (3.33) are correct, the ratio in the last column of Table~\ref{tab:growth} passes through unity at $\alpha\simeq0.16$, being about 0.41 at small $\alpha$ and reaching 20 at $\alpha=0.30$. The absolute value of $\omega_{I}$ given by Eq.~(3.28) is therefore trustworthy only up to a factor of $O(1)$--$O(10)$. Three sources contribute to this discrepancy. First, the cut-off prescription of Eq.~(3.9) is phenomenological: by Eq.~(3.11) it imposes the real logarithmic derivative $+\kappa_{\max}$ at $r_{*}^{\max}$, whereas that of the true solution is fixed by the complex one-way condition of Eq.~(2.32), and the two agree only when the barrier is thick, that is, at small $\alpha$, where $\kappa_{\max}$ is large and the tunnelling weak. Second, the picture of a bound state with weak tunnelling breaks down at large $\alpha$: by Eq.~(3.19), the barrier height $0.176M^{-2}$ of the reference case is already comparable to $\omega^{2}=0.0878M^{-2}$, while the binding energy is only $2.2\times10^{-3}M^{-2}$; as $\alpha$ is increased further, $\omega^{2}$ approaches $V_{\rm eff}(r_{\max})$ and the truncation of Eq.~(3.9) loses its meaning. Third, Eq.~(3.29) is itself an $\alpha\ll1$ result, so that the ratio at $\alpha\gtrsim0.2$ contains the errors of both methods and cannot be assigned to either. Taken together, these considerations give $\alpha\lesssim0.15$ as the range within which the scaling law and the order of magnitude are reliable; a quantitative growth timescale should be taken from Eq.~(3.29) or from the continued-fraction method.

\par
A complex-frequency two-sided shooting was also attempted here in order to obtain $\omega_{I}$ rigorously. The integration was started from Eq.~(2.32) on the horizon side and from Eq.~(2.29), with its Coulomb correction $r^{\nu}e^{-\kappa r}$, at infinity, both solutions being integrated to a matching point $r_{m}$ inside the well and required to have equal logarithmic derivatives, that is,
\begin{equation}
W(\omega)\equiv\frac{\psi_{H}'}{\psi_{H}}\bigg|_{r_{m}}-\frac{\psi_{\infty}'}{\psi_{\infty}}\bigg|_{r_{m}}=0 ,
\label{eq-3.34}
\end{equation}
after which a complex Newton root search was performed on Eq.~(3.34). The method failed in double precision, and the reason is quantitative: $\omega_{I}/\omega_{R}\sim2\times10^{-6}$ for the reference case, while $\psi_{H}$ is amplified by the same factor as it is integrated outwards through the barrier, so that the part of Eq.~(3.34) which carries the information on $\omega_{I}$ is swamped by round-off error. The iteration converged to $M\omega=0.3160+2.61\times10^{-4}i$, whose real part $\omega_{R}=0.316>\mu=0.30$ violates the quasibound-state condition stated below  Eq.~(2.28) and which is therefore a spurious root.

\par
The rigorous procedure is Leaver's continued-fraction method in its massive scalar version. Its starting point is to write the solution of Eq.~(2.19) as
\begin{equation}
R(r)=\big(r-r_{+}\big)^{-i\sigma}\big(r-r_{-}\big)^{i\sigma+\chi-1}e^{-qr}
\sum_{n=0}^{\infty}a_{n}\Big(\frac{r-r_{+}}{r-r_{-}}\Big)^{n} ,
\label{eq-3.35}
\end{equation}
in which the three exponents are fixed by the indicial equations at the two singular points of Eq.~(2.19),
\begin{equation}
q=\sqrt{\mu^{2}-\omega^{2}}\ \ (\mathrm{Re}\,q>0) ,
\qquad
\chi=\frac{\mu^{2}-2\omega^{2}}{q} ,
\qquad
\sigma=\frac{2Mr_{+}\omega-am}{r_{+}-r_{-}} .
\label{eq-3.36}
\end{equation}
The factor $e^{-qr}$ in Eq.~(3.35) ensures decay at infinity, in agreement with Eq.~(2.29), while $(r-r_{+})^{-i\sigma}$ implements the one-way horizon condition of Eq.~(2.32); it may be noted that $\sigma$ in Eq.~(3.36) is proportional to $\varpi(r_{+})$ of Eq.~(2.30). Substituting Eq.~(3.35) into Eq.~(2.19) and collecting powers of $z=(r-r_{+})/(r-r_{-})$, a three-term recurrence $\alpha_{n}a_{n+1}+\beta_{n}a_{n}+\gamma_{n}a_{n-1}=0$ is obtained. A quasibound state requires $\{a_{n}\}$ to be the minimal solution of that recurrence, a condition equivalent to the continued-fraction equation
\begin{equation}
\beta_{0}-\cfrac{\alpha_{0}\gamma_{1}}{\beta_{1}-\cfrac{\alpha_{1}\gamma_{2}}{\beta_{2}-\cfrac{\alpha_{2}\gamma_{3}}{\beta_{3}-\cdots}}}=0 ,
\label{eq-3.37}
\end{equation}
whose complex roots in $\omega$ give $\omega_{R}+i\omega_{I}$. Three points deserve attention in the implementation. The coefficients $\alpha_{n},\beta_{n},\gamma_{n}$ should be derived on the spot from Eqs.~(3.35) and (3.36) with a computer algebra system rather than transcribed, and should be verified by inserting the solution of the recurrence back into Eq.~(2.19) and confirming that the residual vanishes. Thirty to forty decimal digits are required in Eq.~(3.37) in order to accommodate $\omega_{I}/\omega_{R}\lesssim10^{-8}$. The results may finally be cross-checked against $\omega_{R}$ of Eq.~(3.23) and the scaling law of Eq.~(3.33).

\par
It should be stressed, finally, that the reservations expressed above concern $\omega_{I}$ alone and do not extend to the other results of this section. The frequency $\omega_{R}$ of Eq.~(3.18) is given exactly by the cut-off potential, since, as argued below Eq.~(3.9), $\tilde V_{\rm eff}$ and $V_{\rm eff}$ coincide for $r_{*}>r_{*}^{\max}$, which is precisely where the cloud is localized; the $\alpha^{4}$ scaling of Eq.~(3.24) in Table~\ref{tab:spectrum} confirms this independently. The wave function, and with it the particle number of Eq.~(2.39), the energy of Eq.~(2.45) and the angular momentum of Eq.~(2.42), are unaffected for the same reason. The nonlinear results of Sec.~\ref{sec:4} are built on $\omega_{R}$ and on the wave function, and therefore do not inherit this error; the discussion of superradiant saturation given there uses only the sign and the scaling of $\omega_{I}$, both of which have been confirmed by Table~\ref{tab:sign} and Eq.~(3.32) respectively.

\section{Condensate structure and superradiant saturation}
\label{sec:4}
\subsection{Radial structure from the Gross--Pitaevskii equation}
\label{sec:4.1}
\par
The Gross--Pitaevskii equation (2.55) is solved in this section. The control parameters and the numerical method are established in Sec.~\ref{sec:4.1}, together with the radial structure of the condensate, while its geometrical shape, the applicability of the Thomas--Fermi approximation and the mechanism of superradiant saturation are discussed in Sec.~\ref{sec:4.2}. The reference case is again that of Eq.~(3.1), whose linear solution is $M\omega_{0}=0.296304586912$ from Eq.~(3.18). The corresponding binding energy,
\begin{equation}
\mathcal{B}\equiv\mu^{2}-\omega_{0}^{2}=2.2036\times10^{-3}M^{-2} ,
\label{eq-4.1}
\end{equation}
provides the natural scale against which the rise of the chemical potential is measured. Since $\lambda$ enters Eq.~(2.55) only through the combination $\lambda_{\rm eff}|\psi|^{2}$, it may be eliminated by the rescaling
\begin{equation}
v(r_{*})\equiv\sqrt{\lambda}\,\psi(r_{*}) .
\label{eq-4.2}
\end{equation}
Substituting Eq.~(4.2) into Eq.~(2.55) and multiplying throughout by $\sqrt{\lambda}$, an equation free of $\lambda$ is obtained,
\begin{equation}
-\frac{d^{2}v}{dr_{*}^{2}}+V_{\rm eff}(r;\omega)\,v+\hat\lambda_{\rm eff}(r)\,v^{3}=\omega^{2}v ,
\qquad
\hat\lambda_{\rm eff}\equiv\frac{\lambda_{\rm eff}}{\lambda}=\frac{\Delta(r)\,\mathcal{I}(r)}{(r^{2}+a^{2})^{3}} ,
\label{eq-4.3}
\end{equation}
in which $\mathcal{I}(r)$ is given by Eq.~(2.52). Equation (4.3) shows that the shape of the condensate is independent of the numerical value of $\lambda$, which is consistent with the observation made in the spherically symmetric case that the results depend on $\lambda N$ alone.

\par
The physical quantities of Eqs.~(2.39), (2.45) and (2.42) should accordingly be expressed in the same combination with $\lambda$. Inserting Eq.~(4.2) into Eq.~(2.39), the particle number is
\begin{equation}
\lambda N=4\pi\omega\!\int_{-\infty}^{+\infty}\!v^{2}\,w(r)\,dr_{*} ,
\label{eq-4.4}
\end{equation}
with $w(r)$ given by Eq.~(2.40). In order to simplify the quartic term of Eq.~(2.45), use is made of $|\Phi|^{4}=|\psi|^{4}|S|^{4}/(r^{2}+a^{2})^{2}$ and $\sqrt{-g}=\Sigma\sin\theta$ from Eqs.~(2.13) and (2.22), together with the change of variable of Eq.~(2.6), which gives
\begin{equation}
\int|\Phi|^{4}\sqrt{-g}\,d^{3}x
=2\pi\!\int\!dr\,\frac{|\psi|^{4}}{(r^{2}+a^{2})^{2}}
\underbrace{\int_{0}^{\pi}\!|S|^{4}\Sigma\sin\theta\,d\theta}_{=\;\mathcal{I}(r)}
=2\pi\!\int\!\hat\lambda_{\rm eff}(r)\,|\psi|^{4}\,dr_{*} .
\label{eq-4.5}
\end{equation}
The angular integral that appears here is precisely $\mathcal{I}(r)$ of Eq.~(2.52), because the quartic term has the same structure as the projection integral defined above that equation. Substituting Eq.~(4.5) into Eq.~(2.45) and using Eq.~(4.2), and taking the angular momentum directly from Eq.~(2.42), one obtains
\begin{equation}
\lambda E=\omega\,(\lambda N)-\pi\!\int\hat\lambda_{\rm eff}(r)\,v^{4}\,dr_{*} ,
\qquad
\lambda J_{z}=m\,(\lambda N) .
\label{eq-4.6}
\end{equation}
Equations (4.3), (4.4) and (4.6) show that, once $(a,\alpha,\ell,m,n)$ has been fixed, the whole nonlinear problem is governed by the single physical parameter $\mathcal{N}\equiv\lambda N$. The range $\mathcal{N}\in[10^{-3},10^{4}]$ is scanned in this section.

\par
The dimension of $\mathcal{N}$ requires clarification. The action of Eq.~(2.10) gives $[\Phi]^{2}=\hbar/L^{2}$, where $L$ denotes a length, and substitution into $[\lambda][\Phi]^{4}=[\mu^{2}][\Phi]^{2}$ yields
\begin{equation}
[\lambda]=\frac{L^{-2}}{[\Phi]^{2}}=\frac{1}{\hbar} .
\label{eq-4.7}
\end{equation} Since $[N]=[\hbar]$ by Eq.~(2.37), Eq.~(4.7) gives $[\lambda N]=1$; moreover $[v]^{2}=[\lambda][\psi]^{2}=1$ by Eq.~(4.2), so that $v$ is dimensionless, which is why Eq.~(4.3) can be solved purely numerically. It should be pointed out that $\lambda$ is declared in the spherically symmetric literature to carry the dimension of a length and to equal the scattering length, which disagrees with Eq.~(4.7). The discrepancy originates in a different normalization convention for $\Phi$: if the non-relativistic wave-function normalization $[\Phi]^{2}=L^{-3}\times$ particle number is adopted, then $[\lambda]=L$. Before $\mathcal{N}$ is converted into a definite scattering length and particle number $N_{\rm count}$, the normalization of $\Phi$ must therefore be fixed and matched to the non-relativistic Gross--Pitaevskii equation. All conclusions of this section are stated in terms of the dimensionless $\mathcal{N}$ and do not depend on that conversion. As an order-of-magnitude check, let $\hat\lambda$ denote the dimensionless coupling for which $\mathcal{N}=\hat\lambda N_{\rm count}$ holds; the energy fraction of the cloud is then
\begin{equation}
\frac{E_{\rm cloud}}{Mc^{2}}=\big(M\omega\big)N_{\rm count}\left(\frac{l_{P}}{M}\right)^{2}
=\frac{\big(M\omega\big)\mathcal{N}}{\hat\lambda}\left(\frac{l_{P}}{M}\right)^{2} ,
\label{eq-4.8}
\end{equation}
where $l_{P}$ is the Planck length. Since $(l_{P}/M)^{2}\sim10^{-76}$ for $M=M_{\odot}$ and $\sim10^{-96}$ for $M=10^{10}M_{\odot}$, the test-field condition $E_{\rm cloud}/Mc^{2}\ll1$ stated below Eq.~(2.12) is satisfied by an enormous margin over the entire range of $\mathcal{N}$ scanned here, unless $\hat\lambda$ is extremely small.

\par
The solution of Eq.~(4.3) is discussed next. A natural but unworkable approach is to fix $\omega$ and iterate directly, since $v\equiv0$ is always a solution of Eq.~(4.3) while its Jacobian
\begin{equation}
\mathbb{J}\big|_{v=0}=-\frac{d^{2}}{dr_{*}^{2}}+V_{\rm eff}-\omega^{2}
\label{eq-4.9}
\end{equation}
is nonsingular for $\omega\neq\omega_{0}$. By Eq.~(4.9), $v=0$ is therefore an attracting fixed point of the Newton iteration. When continuation in increasing $\mathcal{N}$ was attempted without rescaling the starting guess, the iteration was indeed found to return a sequence of spurious solutions with $\max|v|\sim10^{-18},10^{-28},10^{-38}$. Before a workable scheme is given, the direction of the bifurcation is established. Multiplying Eq.~(4.3) by the linear mode $\psi_{0}$, normalized by $\int\psi_{0}^{2}dr_{*}=1$, integrating, and using the self-adjointness of the linear operator, $\int\psi_{0}[-\partial_{r_{*}}^{2}+V_{\rm eff}]v=\omega_{0}^{2}\int\psi_{0}v$, one obtains
\begin{equation}
\big(\omega^{2}-\omega_{0}^{2}\big)\int v\,\psi_{0}\,dr_{*}=\int\hat\lambda_{\rm eff}\,v^{3}\psi_{0}\,dr_{*} .
\label{eq-4.10}
\end{equation}
In Eq.~(4.3) one has $\hat\lambda_{\rm eff}>0$, and the ground state is nodeless so that $v>0$; the right-hand side of Eq.~(4.10) is therefore positive, whence $\omega>\omega_{0}$. A repulsive self-interaction raises the chemical potential. In the small-amplitude limit, setting $v\simeq A\psi_{0}$, Eq.~(4.10) gives $\omega^{2}-\omega_{0}^{2}=A^{2}\!\int\hat\lambda_{\rm eff}\psi_{0}^{4}$, while Eq.~(4.4) with $w\simeq1$ gives $A^{2}=\mathcal{N}/(4\pi\omega)$. Eliminating $A$ between the two yields
\begin{equation}
\omega^{2}-\omega_{0}^{2}\simeq\frac{\mathcal{N}}{4\pi\omega}\int\hat\lambda_{\rm eff}(r)\,\psi_{0}^{4}\,dr_{*}
\qquad(\mathcal{N}\ll1) .
\label{eq-4.11}
\end{equation}

\par
In order to circumvent the trivial-solution trap described by Eq.~(4.9), $(v,\,E\equiv\omega^{2})$ are taken as the unknowns and a normalization constraint is appended. Equation (4.3) is accordingly written as
\begin{equation}
F(v,E)\equiv-\frac{d^{2}v}{dr_{*}^{2}}+\big(V_{\rm eff}-E\big)v+\hat\lambda_{\rm eff}v^{3}=0 ,
\label{eq-4.12}
\end{equation}
and Eq.~(4.4) supplies the additional condition
\begin{equation}
G(v)\equiv\int v^{2}w\,dr_{*}-\mathcal{N}_{*}=0 ,
\qquad \mathcal{N}_{*}\equiv\frac{\mathcal{N}}{4\pi\omega} .
\label{eq-4.13}
\end{equation}
Equations (4.12) and (4.13) are discretized on the grid of Eq.~(3.15) and with the cut-off potential of Eq.~(3.9). Noting that $\partial F/\partial E=-v$, the bordered Newton system is obtained,
\begin{equation}
\begin{pmatrix}
\mathbb{J} & -v\\[2pt]
2h\,(w\odot v)^{\!\top} & 0
\end{pmatrix}
\begin{pmatrix}\delta v\\[2pt]\delta E\end{pmatrix}
=-\begin{pmatrix}F\\[2pt]G\end{pmatrix} ,
\label{eq-4.14}
\end{equation}
where $\odot$ denotes the elementwise product and the principal block of the Jacobian is the tridiagonal matrix
\begin{equation}
\mathbb{J}=\mathrm{tridiag}\Big(-\frac{1}{h^{2}},\;\frac{2}{h^{2}}+V_{i}-E+3\hat\lambda_{i}v_{i}^{2},\;-\frac{1}{h^{2}}\Big) .
\label{eq-4.15}
\end{equation}
No dense matrix need be assembled for Eq.~(4.14). Two tridiagonal systems $\mathbb{J}y_{1}=-F$ and $\mathbb{J}y_{2}=v$ are solved, and elimination of the second row of Eq.~(4.14) then gives
\begin{equation}
\delta E=\frac{-G-\big(2h\,w\odot v\big)\!\cdot\!y_{1}}{\big(2h\,w\odot v\big)\!\cdot\!y_{2}} ,
\qquad
\delta v=y_{1}+\delta E\,y_{2} ,
\label{eq-4.16}
\end{equation}
so that only two $O(N)$ tridiagonal solves are required per iteration. The step length is set by a damped line search, the residual norm being required to decrease monotonically and the step halved otherwise. The convergence criterion is taken in the relative form
\begin{equation}
\max\big\{\|F\|_{\infty},\,|G|\big\}<\epsilon_{\rm tol}\max\big\{1,\,\|v\|_{\infty}^{3}\big\} ,
\qquad \epsilon_{\rm tol}=10^{-12} ,
\label{eq-4.17}
\end{equation}
in which the factor $\|v\|_{\infty}^{3}$ is necessary, since the nonlinear term of Eq.~(4.12) grows as $v^{3}$ and an absolute criterion cannot be met at large amplitude. The dependence of $V_{\rm eff}$ on $\omega=\sqrt{E}$ still requires the outer loop of Eq.~(3.16), which converges in one or two passes. The solutions are obtained along a sequence of increasing $\mathcal{N}$, each step being started from the $(v,E)$ of the previous one; zero to one Newton iteration per step was needed over the whole scan, with a residual never exceeding $2\times10^{-12}$.

\par
The results are collected in Table~\ref{tab:scan}. Three trends are consistent with $\omega>\omega_{0}$ as obtained from Eq.~(4.10). The chemical potential rises monotonically towards $\mu=0.30M^{-1}$, the third column increasing from $10^{-6}$ to 0.696, so that the binding energy of Eq.~(4.1) is progressively filled. The peak of the cloud moves from $39.4M$ to $122.8M$, that is, the cloud expands under the repulsion. The ratio $E/(\hbar\omega N)$ tends to unity, its departure from unity being $\lambda\!\int\!|\Phi|^{4}\sqrt{-g}/(2\hbar\omega N)$ according to Eq.~(2.45); numerically this ranges from $4\times10^{-9}$ at $\mathcal{N}=10^{-3}$ to $1.9\times10^{-3}$ at $\mathcal{N}=10^{4}$, growing monotonically with $\mathcal{N}$ and of the correct order, which confirms the exact relation (2.45) directly.
\begin{table*}[htbp]
\caption{Solutions of the Gross--Pitaevskii equation (\ref{eq-4.3}) for the reference case. The third column gives the rise of the chemical potential in units of the binding energy (\ref{eq-4.1}); the last tests the exact relation (\ref{eq-2.45}).}
\label{tab:scan}
\centering
\begin{tabular}{cccccc}
\hline\hline
$\mathcal{N}=\lambda N$ & $M\omega$ & $(\omega^{2}-\omega_{0}^{2})/\mathcal{B}$ &
$r_{\rm peak}/M$ & $M\lambda E$ & $E/(\hbar\omega N)$ \\
\hline
$10^{-3}$ & 0.29630458933 & $1\times10^{-6}$ & 39.44 & $2.96305\times10^{-4}$ & 0.999999996 \\
$10^{-2}$ & 0.29630461107 & $6\times10^{-6}$ & 39.44 & $2.96305\times10^{-3}$ & 0.999999959 \\
$10^{-1}$ & 0.29630482850 & $6.5\times10^{-5}$ & 39.44 & $2.96305\times10^{-2}$ & 0.999999592 \\
$1$ & 0.29630700027 & $6.49\times10^{-4}$ & 39.46 & $2.96306\times10^{-1}$ & 0.999995930 \\
$10$ & 0.29632846983 & $6.42\times10^{-3}$ & 39.66 & $2.96317$ & 0.999959864 \\
$10^{2}$ & 0.29652136231 & $5.83\times10^{-2}$ & 41.49 & $2.96417\times10^{1}$ & 0.999647599 \\
$10^{3}$ & 0.29749501965 & $3.208\times10^{-1}$ & 55.38 & $2.97030\times10^{2}$ & 0.998437067 \\
$3\times10^{3}$ & 0.29820908417 & $5.138\times10^{-1}$ & 76.18 & $8.92864\times10^{2}$ & 0.998028749 \\
$10^{4}$ & 0.29888078403 & $6.958\times10^{-1}$ & 122.79 & $2.98325\times10^{3}$ & 0.998141015 \\
\hline\hline
\end{tabular}
\end{table*}

\par
\begin{figure*}[htbp]
\centering
\includegraphics[width=\textwidth]{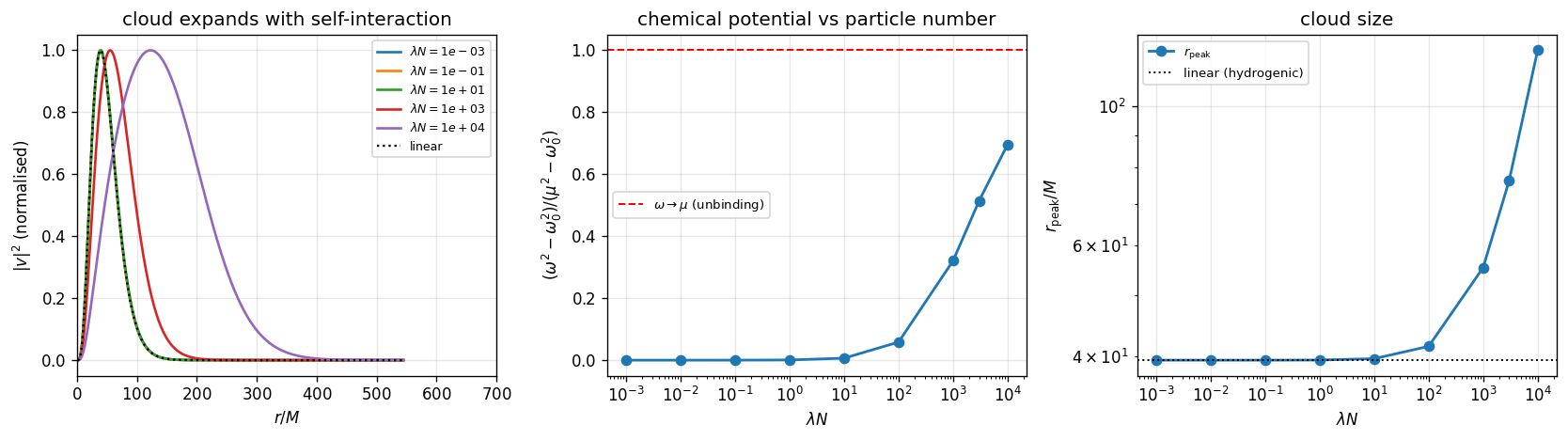}
\caption{Solutions of the Gross--Pitaevskii equation for the reference case. (a) Normalized $|v|^{2}$ against $r/M$ for $\mathcal{N}=10^{-3},10^{-1},10,10^{3},10^{4}$, the linear solution being dotted; the cloud expands as the repulsion is increased. (b) Rise of the chemical potential, in units of the binding energy, against $\mathcal{N}$; the dashed line marks the unbinding limit. (c) Peak radius against $\mathcal{N}$, the dotted line giving the value for the linear solution.}
\label{fig:9}
\end{figure*}

The small-amplitude end of Table~\ref{tab:scan} may be compared quantitatively with Eq.~(4.11). For the reference case the linear mode satisfies
\begin{equation}
\int\hat\lambda_{\rm eff}\,\psi_{0}^{4}\,dr_{*}=5.32\times10^{-6}M^{-1} ,
\label{eq-4.18}
\end{equation}
and substitution of Eq.~(4.18) together with $\mathcal{N}=1$ and $M\omega=0.2963$ into Eq.~(4.11), normalized by $\mathcal{B}$ of Eq.~(4.1), gives
\begin{equation}
\omega^{2}-\omega_{0}^{2}\simeq\frac{5.32\times10^{-6}}{4\pi\times0.2963}=1.429\times10^{-6}M^{-2} ,
\qquad
\frac{\omega^{2}-\omega_{0}^{2}}{\mathcal{B}}=6.48\times10^{-4} ,
\label{eq-4.19}
\end{equation}
whereas the row $\mathcal{N}=1$ of Table~\ref{tab:scan} gives $6.49\times10^{-4}$; the two agree to three significant figures. The four rows from $\mathcal{N}=10^{-2}$ to $10$ also obey the linear scaling of Eq.~(4.11), successive ratios being close to ten. The linear relation fails once $\mathcal{N}\gtrsim10^{2}$, a tenfold increase of $\mathcal{N}$ raising the third column by a factor of only 5.5, because the expansion of the cloud makes the integral $\int\hat\lambda_{\rm eff}v^{4}$ of Eq.~(4.11) grow more slowly than $v^{4}$. The parameter space may accordingly be divided into three regimes,
\begin{equation}
\begin{cases}
\mathcal{N}\lesssim10 : & \text{quasilinear, }\;\omega-\omega_{0}\lesssim10^{-5}\mu ,\\[3pt]
10^{2}\lesssim\mathcal{N}\lesssim10^{3} : & \text{transitional, }\;\mathcal{B}\;\text{filled by }5\%\text{--}30\% ,\\[3pt]
\mathcal{N}\gtrsim10^{3} : & \text{Thomas--Fermi, }\;\omega\to\mu ,\;r_{\rm peak}\sim\mathcal{N}^{1/3} .
\end{cases}
\label{eq-4.20}
\end{equation}
In the first of these the shape of the cloud is indistinguishable from that of the linear solution. Equation (4.20) delimits the range over which the self-interaction matters: it is shown in Sec.~\ref{sec:4.2} that the Thomas--Fermi approximation is usable only in the third regime, and that the identification of the saturation mechanism likewise depends on which regime is realized. The family of solutions of Eq.~(4.3) and its evolution with $\mathcal{N}$ are shown in Fig.~\ref{fig:9}. The upper limit imposed by the grid should finally be noted. For $\mathcal{N}>10^{4}$ the cloud extends beyond the outer boundary $556M$ of Eq.~(3.15) and the solution begins to be contaminated by the Dirichlet condition; at $\mathcal{N}=3\times10^{4}$ one finds $r_{\rm peak}=231M$ while the outer edge of the support has already reached about $713M$. The reliable range of this work is therefore $\mathcal{N}\lesssim10^{4}$, and an extension to larger $\mathcal{N}$ requires the grid to be enlarged as $r_{\rm peak}\sim\mathcal{N}^{1/3}$ at fixed $h$.

\subsection{Toroidal shape, Thomas--Fermi accuracy, and saturation}
\label{sec:4.2}
\par
The angular shape of the condensate cannot be read off directly from the single-mode solution of Sec.~\ref{sec:4.1}, since that projection has already fixed the angular structure to be $S_{\ell m}$. The Thomas--Fermi approximation is therefore applied directly to the unprojected two-dimensional equation (2.51), the two derivative terms of which are discarded, leaving
\begin{equation}
\Big[\frac{\varpi^{2}}{\Delta}+2am\omega-\frac{m^{2}}{\sin^{2}\theta}-a^{2}\omega^{2}\sin^{2}\theta\Big]\Psi
=\big(\mu^{2}+\lambda|\Psi|^{2}\big)\Sigma\,\Psi .
\label{eq-4.21}
\end{equation}
Equation (4.21) is algebraic in $|\Psi|^{2}$. Solving it and discarding the regions in which the result is negative, the two-dimensional Thomas--Fermi density is obtained,
\begin{equation}
\lambda\big|\Psi(r,\theta)\big|^{2}=\Big[\mathcal{U}_{\rm TF}(r,\theta)\Big]_{+} ,
\quad
\mathcal{U}_{\rm TF}=\frac{1}{\Sigma}\Big[\frac{\varpi^{2}}{\Delta}+2am\omega
-\frac{m^{2}}{\sin^{2}\theta}-a^{2}\omega^{2}\sin^{2}\theta\Big]-\mu^{2} ,
\label{eq-4.22}
\end{equation}
where $[\,\cdot\,]_{+}\equiv\max(\cdot,0)$. The advantage of Eq.~(4.22) is that no single-mode projection is required, so that the angular deformation caused by the self-interaction is automatically included.

\par
\begin{figure*}[htbp]
\centering
\includegraphics[width=\textwidth]{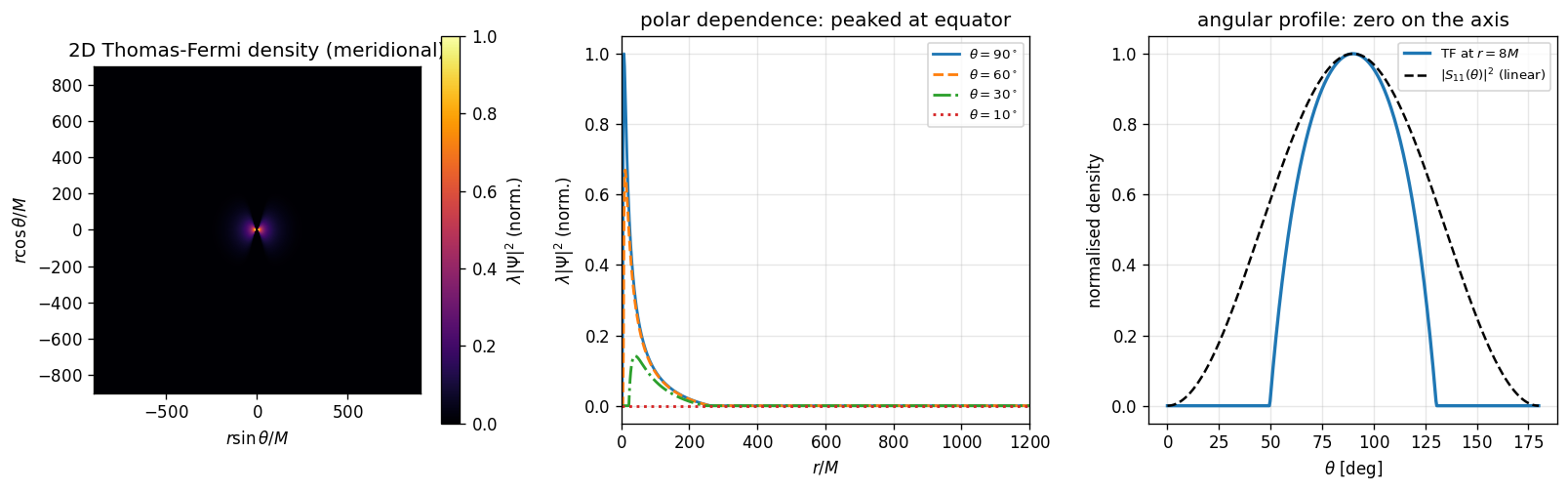}
\caption{Toroidal condensate obtained from the two-dimensional Thomas--Fermi density at $\mathcal{N}=10^{4}$. (a) Density in the meridional plane $(x,z)=(r\sin\theta,\,r\cos\theta)$; the black disc is the horizon and the density vanishes identically on the rotation axis. (b) Radial profiles at $\theta=90^{\circ},60^{\circ},30^{\circ},10^{\circ}$. (c) Angular profile at the density peak $r=8.3M$, compared with the angular function of the linear solution.}
\label{fig:10}
\end{figure*}

The shape of the condensate is decided by the competition between two terms of Eq.~(4.22). The centrifugal barrier $-m^{2}/(\Sigma\sin^{2}\theta)$ diverges to $-\infty$ on the rotation axis, $\theta\to0,\pi$, so that
\begin{equation}
\lambda|\Psi|^{2}\big|_{\theta\to0,\pi}\equiv0
\qquad(m\neq0) ,
\label{eq-4.23}
\end{equation}
that is, the density vanishes identically on the axis rather than merely being small there. The radial attraction $\varpi^{2}/(\Delta\Sigma)$, on the other hand, is largest in the equatorial plane, since $\Sigma=r^{2}+a^{2}\cos^{2}\theta$ of Eq.~(2.2) attains its minimum $r^{2}$ there. Taken together with Eq.~(4.23), these two features produce a ring in the equatorial plane. For the reference case at $\mathcal{N}=10^{4}$, for which Table~\ref{tab:scan} gives $M\omega=0.298881$, the density peak is found at $r=8.26M$ and $\theta=89.82^{\circ}$, the equatorial support is $r\in[4.7M,262.6M]$, and the maximum density near the axis, at $\theta=3^{\circ}$, is strictly zero. By way of contrast, $a=0$ and $m=0$ reduce Eq.~(4.22) in the spherically symmetric case to an expression depending on $r$ alone, and the condensate then occupies a spherical shell. Rotation has thus turned the shell into a torus. This structure is shown in Fig.~\ref{fig:10}.

\par
The numerical results of Eq.~(4.22) may be checked independently by a far-region expansion. In the equatorial plane $\Sigma=r^{2}$, and at large $r$ one has $\varpi\simeq\omega r^{2}$ and $\Delta\simeq r^{2}(1-2M/r)$, so that
\begin{equation}
\frac{\varpi^{2}}{\Delta\Sigma}\simeq\omega^{2}\Big(1+\frac{2M}{r}\Big) ,
\qquad
\mathcal{U}_{\rm TF}\simeq\big(\omega^{2}-\mu^{2}\big)+\frac{2M\omega^{2}}{r}-\frac{m^{2}}{r^{2}} ,
\label{eq-4.24}
\end{equation}
which is a Coulomb attraction plus a centrifugal barrier in the Newtonian limit, of the same structure as Eq.~(3.20). Setting $\mathcal{U}_{\rm TF}=0$ in Eq.~(4.24) determines the inner and outer edges of the density. Substituting $M\omega=0.298881$, $\alpha=0.3$ and $m=1$, for which $\mu^{2}-\omega^{2}=6.699\times10^{-4}M^{-2}$ and $\omega^{2}=0.089330M^{-2}$, a quadratic equation with two roots is obtained,
\begin{equation}
6.699\times10^{-4}r^{2}-0.178660\,r+1=0
\quad\Longrightarrow\quad
r_{1}=5.72M ,\qquad r_{2}=260.9M .
\label{eq-4.25}
\end{equation}
Compared with the numerical support $[4.7M,262.6M]$ quoted above, the outer edge of Eq.~(4.25) agrees to 0.7\% while the inner edge deviates by about 20\%, the latter being caused by relativistic corrections near the horizon. Setting the derivative of Eq.~(4.24) to zero likewise fixes the position of the peak,
\begin{equation}
-\frac{2M\omega^{2}}{r^{2}}+\frac{2m^{2}}{r^{3}}=0
\quad\Longrightarrow\quad
r_{\rm peak}^{\rm TF}=\frac{m^{2}}{M\omega^{2}}=11.19M ,
\label{eq-4.26}
\end{equation}
whereas the numerical value is $8.26M$, the relativistic corrections having moved the peak inwards by about 26\%. It may be noted that Eq.~(4.26) agrees closely with $r_{B}=11.11M$ of Eq.~(3.22), so that the Thomas--Fermi density peaks at the gravitational Bohr radius. Two pathologies of Eq.~(4.22) must further be dealt with. First, $\varpi^{2}/\Delta\to\infty$ makes the density diverge as $r\to r_{+}$; this is a failure of the Thomas--Fermi approximation in the region where the kinetic energy dominates, and the prescription adopted here is to retain only the connected region outside the barrier, that is $r>r_{\max}=2.25M$ from Eq.~(3.19). Second, the derivative of $|\Psi|^{2}$ diverges at the edges $r_{1}$ and $r_{2}$ of Eq.~(4.25), so that the kinetic term diverges when Eq.~(4.22) is inserted back into the original equation; the total energy nevertheless remains finite, and Thomas--Fermi estimates of integrated quantities are therefore still usable. It should be noted that the rigorous justification of the Thomas--Fermi approximation covers only harmonic-oscillator-like external potentials and does not extend to $V_{\rm eff}$ of Eq.~(2.27), which is why the quantitative error curves given below are necessary.

\par
Four distinct objects must be distinguished before Eq.~(4.22) can be compared quantitatively with the numerical solution of Sec.~\ref{sec:4.1}. The quantity $|v|^{2}$ of Eq.~(4.3) contains a Jacobian and is not a density; the radial part of the physical number density is $|v|^{2}/(r^{2}+a^{2})$; discarding the kinetic term of Eq.~(4.3) gives a projected one-dimensional Thomas--Fermi density in the same variable as $|v|^{2}$,
\begin{equation}
\lambda|\psi|^{2}_{\rm TF,1D}=\frac{\omega^{2}-V_{\rm eff}(r)}{\hat\lambda_{\rm eff}(r)}
\qquad\big(\omega^{2}>V_{\rm eff}\;\text{and}\;r>r_{\max}\big) ;
\label{eq-4.27}
\end{equation}
and Eq.~(4.22) gives $\lambda|\Phi|^{2}$ itself. The peak radius of one and the same numerical solution, expressed in these four ways, is listed in Table~\ref{tab:peaks}. Rows one and three, which share a variable, differ by 9\%; rows two and four, which also share a variable, differ by a factor of 2.6; and when the variables are not aligned the peaks may differ by an order of magnitude. The difference between rows two and four has a definite origin. The projection leading to Eq.~(4.27) retains the angular information, through $A_{\ell m}$ and the projection integral of Eq.~(2.52), and discards only the radial kinetic energy, whereas Eq.~(4.22) replaces $A_{\ell m}$ pointwise by $-m^{2}/\sin^{2}\theta$ and thereby discards the angular kinetic energy as well. For $\ell=m=1$ the angular curvature of $S\propto\sin\theta$ is not small, its kinetic contribution being of order $A_{\ell m}-m^{2}=1$, which is comparable to the centrifugal term and cannot be neglected. It follows that the two-dimensional Thomas--Fermi density gives the correct topology and angular profile but a radial scale too small by a factor of 2.6, while the projected one-dimensional density is much more accurate radially but carries no angular information. This distinction does not arise in the spherically symmetric case, where $\ell=0$ and the angular kinetic energy vanishes identically.
\begin{table*}[htbp]
\caption{Peak radius of one and the same Gross--Pitaevskii solution at $\mathcal{N}=10^{4}$, expressed in four variables. Rows one and three share the same variable, as do rows two and four.}
\label{tab:peaks}
\centering
\begin{tabular}{lc}
\hline\hline
quantity & peak radius \\
\hline
$|v|^{2}$ from Eq.~(\ref{eq-4.3}) & $122.79M$ \\
$|v|^{2}/(r^{2}+a^{2})$, number density & $21.76M$ \\
projected Thomas--Fermi, Eq.~(\ref{eq-4.27}) & $133.37M$ \\
two-dimensional Thomas--Fermi, Eq.~(\ref{eq-4.22}) & $8.26M$ \\
\hline\hline
\end{tabular}
\end{table*}

\par
\begin{figure*}[htbp]
\centering
\includegraphics[width=\textwidth]{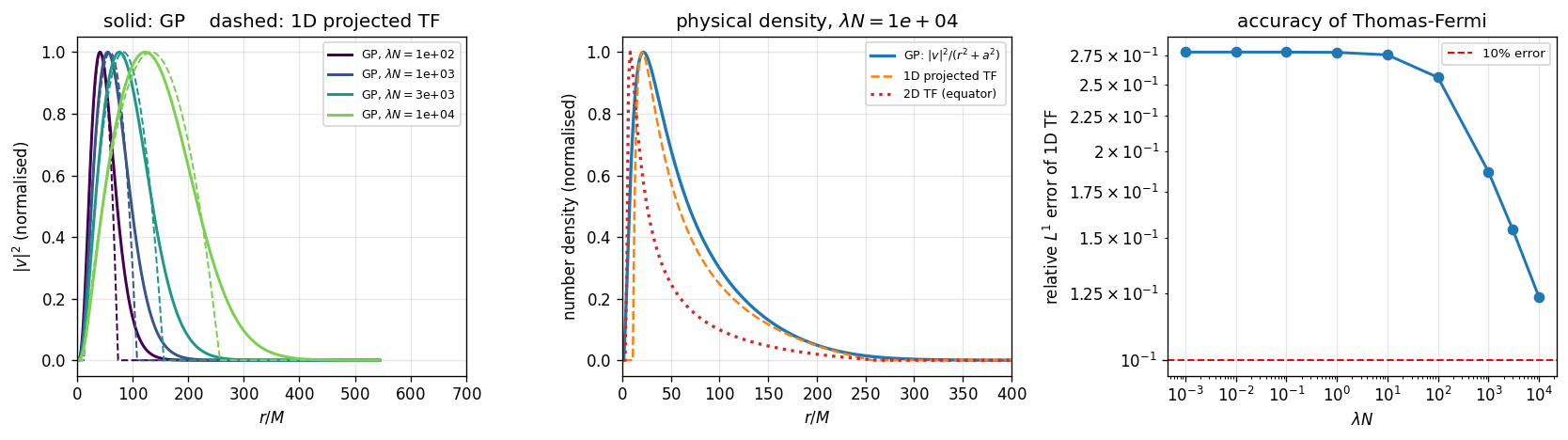}
\caption{Accuracy of the Thomas--Fermi approximation. (a) In the $|v|^{2}$ variable: the numerical solution (solid) against the projected Thomas--Fermi density (dashed) for $\mathcal{N}=10,10^{2},10^{3},10^{4}$. (b) In the number-density variable at $\mathcal{N}=10^{4}$: the numerical solution, the projected Thomas--Fermi density, and the equatorial cut of the two-dimensional one. (c) Relative $L^{1}$ error against $\mathcal{N}$; the dashed line marks 10\%.}
\label{fig:11}
\end{figure*}

With the variables aligned, the departure of Eq.~(4.27) from the numerical solution is measured by the relative $L^{1}$ error of the peak-normalized profiles,
\begin{equation}
\mathcal{E}=\frac{\displaystyle\int\Big|\frac{v^{2}}{\max v^{2}}-\frac{v_{\rm TF}^{2}}{\max v_{\rm TF}^{2}}\Big|dr_{*}}
{\displaystyle\int\frac{v^{2}}{\max v^{2}}\,dr_{*}} ,
\label{eq-4.28}
\end{equation}
and the results are given in Table~\ref{tab:tferr}. The error decreases monotonically with $\mathcal{N}$, from 0.278 to 0.123. Within the quasilinear regime $\mathcal{N}\lesssim10$ of Eq.~(4.20), however, it remains stuck at 0.278, because the kinetic energy is the dominant term there and is precisely what the Thomas--Fermi approximation discards. Even at $\mathcal{N}=10^{4}$ the error is still 12\%. This residual is caused mainly by the hard cut-off of Eq.~(4.22) at the outer edge $r_{2}$, where $|\Psi|^{2}$ is brought to zero with a finite slope, whereas Eq.~(2.29) shows that the true solution has an exponentially decaying tail there; the discrepancy does not disappear as $\mathcal{N}$ is increased. Owing to the grid limitation described below Eq.~(4.20), no threshold for $\mathcal{E}<10\%$ can be given here; extrapolation of the trend suggests $\mathcal{N}\sim10^{5}$. The comparison of Eqs.~(4.27) and (4.22) with the numerical solution is shown in Fig.~\ref{fig:11}.
\begin{table*}[htbp]
\caption{Relative $L^{1}$ error (\ref{eq-4.28}) of the projected Thomas--Fermi density with respect to the numerical solution, as a function of the control parameter.}
\label{tab:tferr}
\centering
\begin{tabular}{cccccccccc}
\hline\hline
$\mathcal{N}$ & $10^{-3}$ & $10^{-2}$ & $10^{-1}$ & $1$ & $10$ & $10^{2}$ &
$10^{3}$ & $3\times10^{3}$ & $10^{4}$ \\
\hline
$\mathcal{E}$ & 0.278 & 0.278 & 0.278 & 0.278 & 0.276 & 0.256 & 0.187 & 0.155 & 0.123 \\
\hline\hline
\end{tabular}
\end{table*}

\par
The saturation of superradiance is discussed last. Substituting $\omega=\omega(\mathcal{N})$ from Sec.~\ref{sec:4.1} and $\omega_{I}=\omega_{I}(\omega)$ from Sec.~\ref{sec:3.2} into the relation $\dot N=2\omega_{I}N$ given above Eq.~(2.48), the evolution equation of the cloud is obtained,
\begin{equation}
\frac{d\mathcal{N}}{dt}=2\,\omega_{I}\big(\omega(\mathcal{N})\big)\,\mathcal{N} .
\label{eq-4.29}
\end{equation}
A saturated state of Eq.~(4.29) requires $\dot{\mathcal{N}}=0$, hence $\omega_{I}=0$ by Eq.~(2.48), and therefore
\begin{equation}
\omega\big(\mathcal{N}_{\rm sat}\big)=m\,\Omega_{H} .
\label{eq-4.30}
\end{equation}
Since $\omega>\omega_{0}$ by Eq.~(4.10), the growth of the cloud does raise $\omega$ towards $m\Omega_{H}$, which is the saturation mechanism anticipated below Eq.~(2.57). That rise is not unbounded, however. The quasibound-state condition stated below Eq.~(2.28) requires $\omega<\mu$, and the monotonic approach of the third column of Table~\ref{tab:scan} to unity shows precisely that $\omega\to\mu$ is the asymptotic limit as $\mathcal{N}\to\infty$; $\omega$ cannot cross $\mu$, since crossing it means unbinding. A necessary condition for Eq.~(4.30) to have a solution is therefore $m\Omega_{H}<\mu$. Combining this with the condition $\omega_{0}<m\Omega_{H}$ for superradiance to occur, that is, with the criterion below Eq.~(2.31) applied to the linear mode, and substituting the spectrum of Eq.~(3.23), one obtains
\begin{equation}
\mu\Big(1-\frac{\alpha^{2}}{2n_{p}^{2}}\Big)\;<\;m\Omega_{H}\;<\;\mu .
\label{eq-4.31}
\end{equation}
Equation (4.31) confines $m\Omega_{H}$ to an interval of width $\mu-\omega_{0}=\mu\alpha^{2}/(2n_{p}^{2})$, whose relative width is
\begin{equation}
\frac{\Delta(m\Omega_{H})}{\mu}=\frac{\alpha^{2}}{2n_{p}^{2}} .
\label{eq-4.32}
\end{equation}
For the reference values $\alpha=0.3$ and $n_{p}=2$, Eq.~(4.32) gives $1.1\times10^{-2}$, so that the window is only about 1\% wide. Over most of the parameter space the self-interaction is therefore unable to switch off superradiance by itself, and saturation must come from the spin-down of the black hole, described by Eq.~(2.46) and with the observational consequences reviewed in Ref.~\cite{Arvanitaki:2010sy}, or from higher-order effects such as energy transfer between modes and bosenova collapse.

\par
The reference case itself lies outside the window of Eq.~(4.31). From Eqs.~(3.1), (3.2) and (3.18),
\begin{equation}
\mu=0.30M^{-1} ,\qquad m\Omega_{H}=0.4338M^{-1} ,\qquad \omega_{0}=0.29630M^{-1} ,
\label{eq-4.33}
\end{equation}
so that $m\Omega_{H}>\mu$ and the right-hand half of Eq.~(4.31) is violated. However large $\mathcal{N}$ is taken, one then has $\omega\to\mu=0.30M^{-1}<m\Omega_{H}=0.4338M^{-1}$, superradiance remains switched on, and Eq.~(4.29) has no saturation fixed point. The margin $m\Omega_{H}-\mu=0.134M^{-1}$ of Eq.~(4.33) is forty times the window width $3.4\times10^{-3}M^{-1}$ of Eq.~(4.32). Even if Eq.~(4.31) were satisfied, a substantial $\mathcal{N}$ would still be needed for $\omega$ to be raised appreciably. From Eq.~(4.11), the control parameter required to raise the chemical potential to the scale of the binding energy (4.1) is
\begin{equation}
\mathcal{N}\sim\frac{4\pi\omega\,\mathcal{B}}{\displaystyle\int\hat\lambda_{\rm eff}\psi_{0}^{4}dr_{*}}
=\frac{4\pi\times0.2963\times2.2036\times10^{-3}}{5.32\times10^{-6}}\approx1.5\times10^{3} ,
\label{eq-4.34}
\end{equation}
which is of the same order as the actual result of Table~\ref{tab:scan}, where a rise of 32\% is reached at $\mathcal{N}=10^{3}$, and is consistent with the third regime of Eq.~(4.20). If, conversely, the value of $\mathcal{N}$ realized astrophysically is only $O(1)$, then by Eq.~(4.19) the chemical potential is raised by no more than $6.5\times10^{-4}$ of the binding energy, the self-interaction has almost no effect on the structure of the cloud, and the cloud is essentially the linear solution of Sec.~\ref{sec:3}. Identifying the saturation mechanism thus hinges on determining the actual value of $\mathcal{N}$, which in turn requires the dimensional bookkeeping described below Eq.~(4.7). It is worth noting that this conclusion echoes a feature of the spherically symmetric case, in which $\lambda$ cancels out completely from the lifetime formula of the cloud. Equations (4.19) and (4.34) provide a quantitative explanation of that feature: at $\mathcal{N}=O(1)$ the self-interaction is a negligible perturbation, and $\mathcal{N}\gtrsim10^{3}$ is required before condensate effects become important.

\section{Conclusions}
\label{sec:5}
\par
In this work, we have performed a comprehensive relativistic and numerical analysis of self-interacting scalar condensates around a Kerr black hole. By focusing on the frame-dragging effect encoded in the cross term $g_{t\varphi}$ of the Boyer--Lindquist metric, we demonstrated that the decaying scalar wigs found in spherically symmetric cases transform into exponentially growing condensates due to rotation. Consequently, the quartic coupling $\lambda$ is no longer just a structural detail; it determines the ultimate fate of the superradiant instability.

\par
By separating the Klein--Gordon equation into spheroidal harmonic and radial parts, we derived a closed-form effective potential. The horizon limit of this potential directly yields the superradiance criterion $0<\omega<m\Omega_{H}$. Because the angular eigenvalue depends on the unknown frequency, we resolved this nonlinear coupling using a self-consistent iteration. To overcome the non-separability of the equations when $\lambda\neq0$, we applied a single-mode projection. This approach yields a one-dimensional Gross--Pitaevskii equation with a geometrically modulated effective coupling $\lambda_{\rm eff}=\lambda\Delta\mathcal{I}(r)/(r^{2}+a^{2})^{3}$. By rescaling the field as $v=\sqrt{\lambda}\,\psi$, we completely absorbed this coupling and isolated the physical effects of the self-interaction into a single dimensionless control parameter, $\mathcal{N}=\lambda N$.

\par
Our analysis reveals that the self-interaction is effectively switched off at the horizon by the geometric factor $\Delta$ and only becomes significant outside the potential barrier. As a result, the strong-field structure of the condensate is substantially modified, while its outer envelope remains essentially unperturbed. This geometric modulation reorganizes the density from the spherical shell seen in static cases into a torus, forcing the density to vanish identically along the rotation axis. Based on the cloud's response to the self-interaction, we identified three distinct regimes: a quasilinear regime ($\mathcal{N}\lesssim10$) where the cloud is indistinguishable from the linear solution; a transitional regime; and a Thomas--Fermi regime ($\mathcal{N}\gtrsim10^{3}$) where the chemical potential approaches $\mu$ and the cloud expands as $\mathcal{N}^{1/3}$. Furthermore, we compared the two-dimensional and the projected Thomas--Fermi densities and found they differ by a factor of 2.6 in their radial scale. This discrepancy quantifies the angular kinetic energy neglected by the 2D approximation---a distinction that does not exist in spherically symmetric models.

\par
Our growth-rate calculations independently reproduce the analytic scaling $\omega_{I}\propto\alpha^{4\ell+5}$ using the conserved-flux formula. However, because the absolute coefficient depends on the specific cut-off prescription, this numerical rate is strictly reliable only for $\alpha\lesssim0.15$. From a methodological standpoint, we successfully bypassed the trivial solution that plagues fixed-frequency iterations by implementing a bordered Newton scheme at a fixed particle number. This allowed us to trace the entire solution family through continuation in $\mathcal{N}$, requiring only two tridiagonal matrix solves per iteration. With this numerical scheme, we successfully recovered the two exact conservation relations, $J_{z}=\hbar mN$ and $E=\hbar\omega N-\frac{\lambda}{2}\!\int|\Phi|^{4}\sqrt{-g}\,d^{3}x$, across six orders of magnitude in the quartic correction. Converting these results into physical units using $\alpha=0.7485\,(M/M_{\odot})(M_{\rm boson}c^{2}/10^{-10}\,\mathrm{eV})$, we find that a $10M_{\odot}$ stellar-mass black hole at $\alpha=0.1$ emits a $650\,\mathrm{Hz}$ signal with an $e$-folding time of roughly 30 days. Conversely, a $10^{7}M_{\odot}$ supermassive black hole produces a $6.5\times10^{-4}\,\mathrm{Hz}$ signal over approximately $8\times10^{4}$ years. Both signals fall squarely within the optimal sensitivity bands of ground-based and space-borne gravitational-wave interferometers, respectively.

\par
Finally, we established a definitive criterion for superradiant saturation: the repulsion-induced frequency shift must reach $m\Omega_{H}$ before hitting the quasibound-state limit $\omega<\mu$. This requirement confines the admissible horizon frequencies to a narrow window of relative width $\alpha^{2}/2n_{p}^{2}$ (about 1\% at $\alpha=0.3$, narrowing as $\alpha^{2}$ for smaller couplings). Consequently, across most of the parameter space, the self-interaction alone cannot quench the superradiant instability. Instead, saturation must be driven by the spin-down of the black hole or by higher-order mode couplings. This finding quantitatively explains why $\lambda$ cancels out in the lifetime formulas of spherically symmetric models. Looking forward, extending this framework to include the continued-fraction method, the full two-dimensional nonlinear geometry, and self-gravitating backgrounds will further cement rotating scalar condensates as premier astrophysical laboratories for probing ultralight dark matter in the extreme strong-field regime.


\acknowledgments

This work is supported by the National Natural Science Foundation of China (Grant No. 12505060, 12375043, 12575069, 12175027, 11875010), the Fund Project of Chongqing Normal University (Grant Number: 24XLB033, 26XLB001), Chongqing Natural Science Foundation General Program (Grant No. CSTB2025NSCQ-GPX1019, and Science and Technology Research Project of Chongqing Municipal Education Commission (Grant Number: KJQN202500563).


\end{document}